%% file: main.tex
\documentclass[floatfix,
 reprint,
 superscriptaddress,
 amsmath,amssymb,
 aps,
 prd,
 nofootinbib,
 longbibliography
]{revtex4-2}

\usepackage{graphicx}
\usepackage{dcolumn}
\usepackage{bm}
\usepackage{siunitx}
\usepackage{physics}
\usepackage{xspace}
\usepackage{mathrsfs}
\usepackage{slashed}
\usepackage{multirow}
\usepackage{hyperref}
\usepackage[capitalize,noabbrev]{cleveref}
\hypersetup{
  colorlinks=true,
  allcolors=blue,
  pdftitle={Probing Long-Lived Particle Production in Muon Decays at the SNS with a Highly Capable Hydrocarbon Detector},
  pdfauthor={PROSPECT Collaboration, M. Hostert, S. Urrea},
}

\newcommand{\nuebar}{\ensuremath{\bar{\nu}_{e}}\xspace}

\newcommand{\Ufive}{\ensuremath{^{235}\text{U}}\xspace}

\newcommand{\HCdet}{HC$^2$\xspace}

\usepackage[mathlines]{lineno}

\begin{document}

\title{Probing Long-Lived Particle Production in Muon Decays at the SNS with a Highly Capable Hydrocarbon Detector}

\input{AuthorListJune2026}

\collaboration{The PROSPECT Collaboration}
\homepage{http://prospect.yale.edu \\ email: prospect.collaboration@gmail.com}

\author{M. Hostert}
\email{matheus-hostert@uiowa.edu}
\affiliation{Department of Physics and Astronomy, University of Iowa, Iowa City, IA 52242, USA}

\author{S. Urrea}
\email{salvador.urrea@ijclab.in2p3.fr}
\affiliation{IJCLab, Pôle Théorie (Bat. 210), CNRS/IN2P3, 91405 Orsay, France}




\begin{abstract}
The Spallation Neutron Source (SNS) at Oak Ridge National Laboratory (ORNL) is a prolific muon producer, making it an ideal location for studying dark sector particles produced in muon decays at rest. 
In this paper, we explore sub-GeV dark particle detection possibilities in a tons-scale, highly capable hydrocarbon scintillator (HC$^2$) detector at the SNS. 
We consider a search for $e^+e^-$ final states produced by decays of long-lived, $\mathcal{O}(10-100)$~MeV axion-like particles and heavy neutral leptons.  
The HC$^2$ technology space, exemplified by the PROSPECT and Mobile Antineutrino Demonstrator detectors, offers strong rejection capabilities for the cosmic ray backgrounds that would normally dominate this search.  
By benchmarking on-surface cosmic ray signatures with data from PROSPECT at ORNL, we generate robust predictions for a multi-year SNS deployment of a range of HC$^2$ detector implementations.  
Results indicate the potential for order-of-magnitude improvements in sensitivity to axion-like particles and heavy neutral leptons in the 10-100 MeV mass regime compared to current global limits.  
We also comment on the neutrino detection possibilities of a HC$^2$ deployment at the SNS.
\end{abstract}

\maketitle

\section{Introduction}

The past decade has seen an evolution and diversification in the hunt for the nature of dark matter.  
In recent years, generations of dedicated low-background detectors sensitive to soft nuclear scatters of weakly interacting massive particles in the cosmic dark matter halo~\cite{EDELWEISS:2011epn,CDMS:2013juh,DarkSide:2014llq,DEAP-3600:2017uua,XENON:2018voc,PICO:2019vsc,LZ:2022lsv,PandaX-4T:2021bab,COSINE-100:2019lgn} have been joined by an increasingly broad array of efforts and technologies probing other potential components of that halo.  Some, such as axion haloscope experiments~\cite{McAllister:2017lkb,ADMX:2018gho,HAYSTAC:2020kwv,Jeong:2020cwz,CAPP:2020utb,GigaBREAD:2025lzq}, have detectors tailored for this purpose, while others, such as boosted dark matter searches, involve the re-purposing of neutrino experiment detectors towards this alternate goal~\cite{Cappiello:2019qsw,Berger:2019ttc,PROSPECT:2021awi,Super-Kamiokande:2022ncz}.  
Other experimentalists have turned their gaze instead to potential terrestrial sources of dark matter and dark particles in general, such as proton beams or nuclear reactors.  
Since these locations host the highest earthly densities of energetic or rare Standard Model particles (pions, kaons, heavier mesons, and GeV-scale neutrinos at proton beams; $\gamma$ rays and MeV-scale neutrinos at nuclear reactors), they are ideal sites for probing new dark sectors that directly or indirectly couple to these particles.

As with the cosmic frontier community, recent dark sector and other beyond-the-Standard-Model (BSM) searches at proton beamlines encompass both purpose-built efforts (such as CCM~\cite{CCM:2021leg,CCM:2021yzc}) as well as parasitic efforts in neutrino experiments built for other physics purposes (such as ND280~\cite{T2K:2019jwa}, Fermilab's LArTPC experiments~\cite{Machado:2019oxb,MicroBooNE:2022ctm,MicroBooNE:2023eef,ICARUS:2024oqb}, COHERENT~\cite{COHERENT:2021pvd}, and others~\cite{MiniBooNE:2017nqe,ArgoNeuT:2019ckq,ArgoNeuT:2022mrm,NOvA:2025ykl}).  
A subset of these efforts proceed by searching for the charged products of a dark sector long-lived particle (LLP) decaying inside a detector far removed (meters to kilometers) from the LLP source.  
The masses of manifested LLPs are primarily limited by the masses of the decaying mesons and secondary particles that produce them, with production of these in turn limited by the kinetic energies of the beam protons.   
Thus, experiments at higher-energy proton sources, such as the T2K-feeding 30~GeV Main Ring at J-PARC
and the 120~GeV NuMI beam at Fermilab can access LLPs with masses near or above the GeV scale, while proton beams with energies at or below 1~GeV, such as the 1.3~GeV Spallation Neutron Source (SNS) at Oak Ridge National Laboratory (ORNL) and the 800~MeV beam at the Los Alamos Neutron Science Center (LANSCE) at Los Alamos National Laboratory, search most effectively below the few~hundred MeV~mass range.  
Within its sensitive mass range, a proton beam's generated LLP density is mostly driven by the strength of the dark sector coupling to normal matter and by the power delivered by the proton beam to its target.  

With these considerations in mind, ORNL's SNS and its 2.0~MW designed beam power and short pulse structure is an ideal host for beam-based searches for LLPs in the 10-100 MeV mass regime.  
The multi-component COHERENT experiment~\cite{Barbeau:2021exu}, originally conceived to measure coherent neutrino-nucleus scattering of SNS-produced neutrinos~\cite{COHERENT:2017ipa,COHERENT:2020iec,COHERENT:2024axu}, has reported its first limits on LLPs produced in $\pi^0$ and $\eta^0$ decays in the SNS target~\cite{COHERENT:2021pvd}.  

More recently, Ref.~\cite{Hostert:2025ffy} presented a broad survey of LLP models that can be accessed via $\mu^+$ decays at rest in the SNS target.  
At both SNS and at J-PARC's Materials and Life Science Facility (MLF) beamline, LLPs from in-target $\mu^+$ decay would be produced many $\mu$s after a proton beam spill, allowing for time-based separation of LLP decay signatures from most beam-related backgrounds.  
As a result, Ref.~\cite{Hostert:2025ffy} shows that COHERENT, as well as the JSNS$^2$ neutrino experiment at J-PARC~\cite{JSNS2:2021hyk}, offer the promise of world-leading sub-GeV LLP sensitivity via searches for their e$^+$e$^-$ final-state decay products.  
However, the extent of this reach is clouded by a lack of knowledge of the ultimate cosmic-induced background level that can be achieved in SNS- and J-PARC-based detectors during this post-beam window.  
The e$^+$e$^-$ LLP decay signature of interest presents a particular challenge, since it does not generate time-coincident detector signatures that are often used to reject backgrounds in beam dump neutrino experiments~\cite{LSND:2001aii,LSND:1997lta,LSND:2001fbw,JSNS2:2024uxo,JSNS2:2024bzf}.  

In this paper, we explore how sub-GeV LLP decay detection possibilities can be expanded via deployment of a highly capable hydrocarbon scintillator detector at the SNS -- an experimental arrangement we refer to hereafter as `HC$^2$.'  
While the SNS already hosts large hydrocarbon detectors for particle physics purposes, such as MARS~\cite{Roecker:2016juf}, these efforts have aimed to survey beam-related backgrounds as a means of improving the particle physics deliverables of other non-hydrocarbon COHERENT detectors~\cite{COHERENT:2021qbu}.
Instead, the HC$^2$ concept at the SNS envisions particle physics use cases more akin to the LSND~\cite{LSND:1996jxj}, KARMEN~\cite{DREXLIN1990490} and JSNS$^2$~\cite{JSNS2:2021hyk} organic scintillator experiments, but performed in a smaller detector featuring new and/or substantially improved detector response capabilities that improve background rejection for isolated electromagnetic activity.  
The PROSPECT reactor antineutrino detector serves as a validating example of the HC$^2$ concept~\cite{prospect_nim}.   While performing sterile neutrino searches~\cite{prospect_osc,prospect_prd,PROSPECT:2024gps} and measurements of \Ufive fission neutrino emissions~\cite{prospect_spec,pros_specfinal} at ORNL's High Flux Isotope Reactor (HFIR), PROSPECT demonstrated excellent background rejection through a combination of high light collection, optical segmentation, $^6$Li doping, pulse shape discrimination, and double-ended light readout.  
Other existing $\mathcal{O}$(t)-mass detectors demonstrate the potential breadth of the HC$^2$ concept space: the Mobile Antineutrino Demonstrator's two segmented plastic scintillator sub-detectors offer powerful and contrasting background rejection techniques~\cite{Haghighat:2018mve,MobileAntineutrinoDemonstratorProject:2025sna}, while the Eos detector aims to explore the benefits of hybrid scintillation-Cerenkov detection technology for various hydrogenous target liquids~\cite{Anderson:2022lbb}.

To demonstrate the LLP detection power of a HC$^2$ campaign at the SNS, we will study the experimental sensitivity of a nominal tons-scale scintillator detector to heavy neutral leptons and axion-like particles produced in $\mu^+$ decays in the SNS.  
To support robust sensitivity estimates, we will use PROSPECT's well-determined detector response and background rejection capabilities as a starting point for this study, while also considering performance variations arising from differing assumed HC$^2$ detector capabilities.  
Our study will search for decays of these LLPs, which will generate isolated 10-100~MeV e$^+$e$^-$ pairs that appear in the detector delayed in time with respect to the SNS proton beam -- a period where cosmic backgrounds are expected to provide the dominant background contribution.  
Cosmic contributions are then rigorously estimated using existing PROSPECT datasets and benchmarked cosmic neutron ($n$) and muon ($\mu^{\pm}$) simulations.  
We show that a tons-scale HC$^2$ campaign can perform low-background searches for e$^+$e$^-$ products of LLP decay at the SNS, improving current leading LLP limits by more than an order of magnitude.
We also explore the relative value of various HC$^2$ design and analysis strategies in reducing backgrounds to BSM signals, specifically demonstrating the power of combining target segmentation and pulse shape discrimination capabilities in a high-light-yield scintillator detector.  

The paper will begin in Section~\ref{sec:flux} by describing the phenomenology of the axion-like particle and heavy neutral lepton signatures targeted in this study.  
The LLP decay signal selection criteria are defined in Section~\ref{sec:select}, with signal-like cosmic background measurement results from PROSPECT data and cosmic Monte Carlo simulations given in Section~\ref{sec:results1}.  
Projected BSM signals and cosmic backgrounds from applying the same selection scheme to a variety of HC$^2$ detector implementations are shown in Section~\ref{sec:results2}.  
Based on these HC$^2$ background results and simulated responses to BSM particle decays described in Section~\ref{sec:response}, LLP search sensitivities are provided in Section~\ref{sec:sens}, with some remarks on neutrino detection in HC$^2$ provided in Section~\ref{sec:neutrino}.  

\section{BSM Particle Production and Decay at the SNS}
\label{sec:flux}

The SNS operates a high-energy pulsed proton beam that generates intense fluxes of charged pions, which stop and decay at rest within the SNS target ($\pi^+ \to \mu^+ \nu_\mu$).  
Consequently, the target acts as a high-intensity, isotropic source of muons ($\mu^+$) which subsequently decay at rest ($\mu^+ \to e^+ \nu_e \bar{\nu}_\mu$) in the beam target.  



\begin{figure*}[t]
    \centering
    \includegraphics[width=0.49\textwidth]{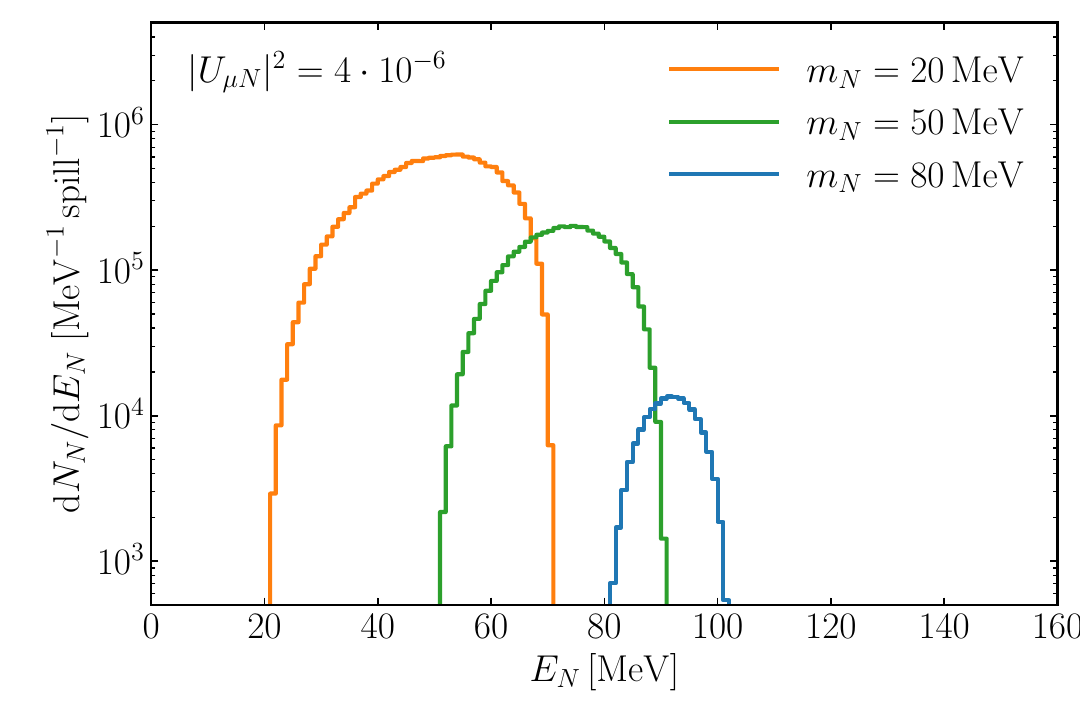}
    \hfill
    \includegraphics[width=0.49\textwidth]{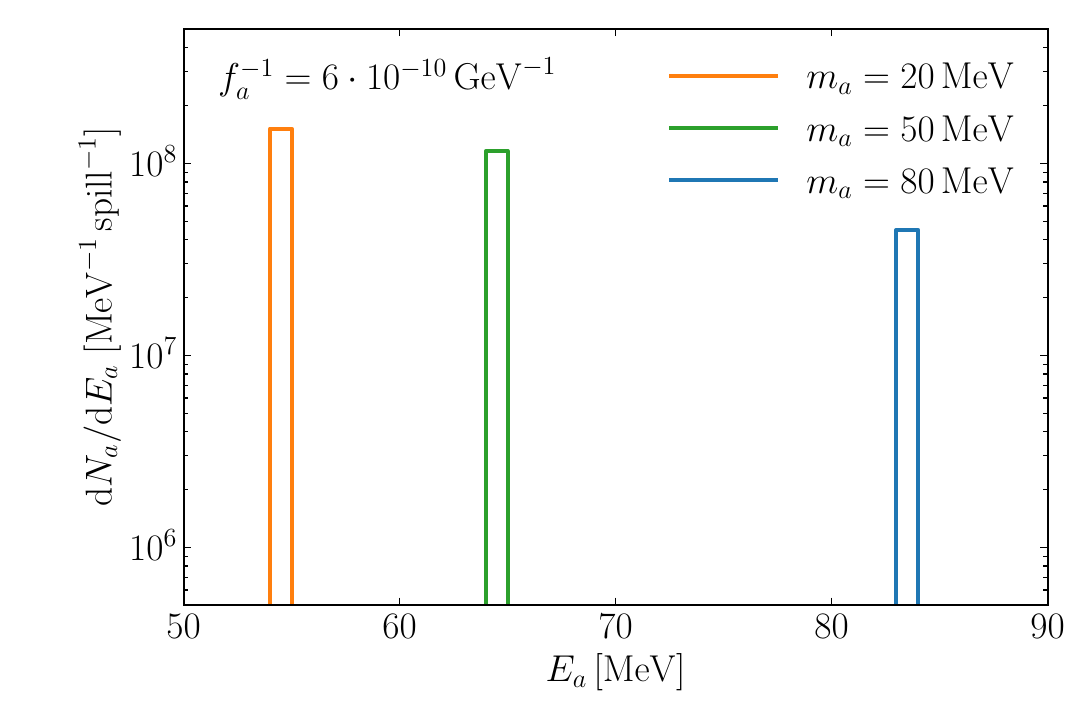}
    \caption{
    Expected flux of HNLs (left panel) and ALPs (right panel) produced in $\mu^{+}$ decay at rest at the SNS. The spectra are shown for three representative masses, $m_{N,a} = 20, 50$ and $80~\mathrm{MeV}$. For both cases we use coupling values near current experimental bounds.
    }
    \label{fig:Flux_HNL}
\end{figure*}

We will focus on two benchmark BSM scenarios produced in the decay of these stopped $\mu^{+}$: heavy neutral leptons (HNLs) and leptophilic axion-like particles (ALPs).
HNLs are a well-motivated extension of the Standard Model that introduces singlet fermions to explain the origin of neutrino masses via the seesaw mechanism~\cite{Minkowski:1977sc,Mohapatra:1979ia,GellMann:1980vs,Yanagida:1979as,Lazarides:1980nt,Mohapatra:1980yp,Schechter:1980gr,Cheng:1980qt,Foot:1988aq}. 
In low-scale seesaw variants~\cite{Mohapatra:1986bd,GonzalezGarcia:1988rw,Wyler:1982dd,Akhmedov:1995ip,Akhmedov:1995vm,Barry:2011wb,Zhang:2011vh} accessible to accelerator experiments, these HNLs ($N$) are pseudo-Dirac particles that mix with the active SM neutrinos. 
The interactions of the $N$ with the weak gauge bosons are suppressed by the mixing matrix elements $U_{\alpha N}$ and are described by the effective low-energy Lagrangian:
\begin{align}
    -\mathscr{L_{\rm HNL}} & \supset \frac{g}{2 c_W} U_{\alpha N}^* \overline{\nu_\alpha} \slashed{Z} P_L N 
    \\\nonumber
    &\quad+ \frac{g}{\sqrt{2}} U_{\alpha N}^* \overline{\ell}_{\alpha} \slashed{W}  P_L N  + \text{ h.c.},
\end{align}
where $\alpha \in \{e, \mu, \tau \}$. While HNLs can be produced in the two-body decays of pions and kaons ($M^+ \to \ell^+ N$), the high-intensity stopped $\mu^{+}$ flux at the SNS offers an alternative production mechanism, particularly for scenarios with dominant muon-flavor mixing ($U_{\mu N}$). In this case, the HNL is produced via the three-body decay $\mu^+ \to e^+ \nu_e N$ for HNL masses below the kinematic threshold ($m_N < m_\mu - m_e$). The differential decay rate is given by~\cite{Ema:2023buz}:
\begin{equation}\label{eq:muon_decay_Umu}
\begin{split}
&\frac{d\Gamma\bigl(\mu \to e\,\nu_e\,N\bigr)}{dE_N}
  ={} \frac{G_F^2\,|U_{\mu N}|^2}{12\,\pi^3}
       \sqrt{E_N^2 - m_N^2} \\[-0.3ex]
     {}&\times\bigl[\,3\,E_N\,(m_\mu^2 + m_N^2)
       - 4\,m_\mu\,E_N^2 - 2\,m_\mu\,m_N^2\bigr].
\end{split}
\end{equation}
Once produced, the HNL can decay invisibly into three neutrinos, into two leptons and a neutrino, or hadronically. In the $m_N$ range targeted in this study (10–100 MeV), the dominant visible decay channel is $N \rightarrow \nu e^+ e^-$, whose differential decay rate can be found in Ref.~\cite{Coloma:2020lgy}.

Axion-Like Particles (ALPs) are light pseudo-Goldstone bosons arising from the breaking of global symmetries. While historically motivated by the Strong CP problem~\cite{Peccei:1977hh,Peccei:1977ur,Wilczek:1977pj,Weinberg:1977ma}, generic MeV scale ALPs have not yet been excluded and may be accessible at spallation sources. We assume the ALP couples to the Standard Model via derivative interactions with a weak-conserving leptophilic current, described by the effective Lagrangian:
\begin{equation}\label{eq:ALPlagrangian}
   \mathscr{L}_{\rm ALP} \supset 
   \frac{1}{2}(\partial_\mu a)(\partial^\mu a)
   - \frac{m_a^2}{2} a^2
   + \frac{\partial_\mu a}{2 f_a} j^\mu_\ell \, .
\end{equation}
with 

\begin{equation}\label{eq:ALPcurrent}
   j^\mu_{\ell} =
   \sum_{i,j}^{e,\mu,\tau}
   c_{ij} \,\left(
   \,\overline{\nu_L^i} \gamma^\mu \nu_L^j-\overline{\ell^i} \gamma^\mu\gamma^5 \ell^j \right)\, .
\end{equation}
where $f_a$ is the ALP decay constant and $c_{ij}$ represent the coupling matrices in flavor space. 

The high-intensity stopped $\mu^{+}$ flux at the SNS provides a unique opportunity to probe Lepton Flavor Violating (LFV) scenarios where off-diagonal couplings (specifically $c_{\mu e}$) are non-zero. This allows for the two-body decay $\mu^+ \to e^+ a$, which dominates over three-body processes and yields a distinct, mono-energetic ALP flux. The total decay rate for $\mu^+ \to e^+ a$ is given by:
\begin{equation}
    \Gamma(\mu^+ \to e^+ a) \simeq |c_{\mu e}|^2  \frac{m_\mu^3}{64 \pi f_a^2} f(r_e, r_a),
\end{equation}where $f(r_e, r_a) = (1 + r_e)^2\left[(1 -r_e)^2 - r_a^2\right]\lambda^{1/2}(1,r_e, r_a)$ with $r_i = m_i/m_\mu$  and $\lambda(x,y,z)=x^2+y^2+z^2-2xy-2xz-2yz$ denoting the Källén function. Then, similarly to HNLs, the ALPs decay via $a \rightarrow e^+ e^-$:
\begin{align}
\Gamma(a \to e^+e^-) = |c_{ee}|^2\frac{m_a m_e^2}{8\pi f_a^2}\left(1-\frac{4 m_e^2}{m_a^2}\right)^{1/2}.
\end{align}

To illustrate the phenomenology of these scenarios, we show in Figure~\ref{fig:Flux_HNL} the fluxes of HNLs and ALPs produced in $\mu^{+}$ decay at rest. 
HNLs are generated via the three-body decay $\mu^+ \to e^+ \nu_e N$, yielding a continuous energy spectrum, whereas LFV ALPs are produced through the two-body decay $\mu^+ \to e^+ a$, resulting in a mono-energetic flux. The overall fluence decreases with increasing mass due to phase-space suppression in the parent $\mu^{+}$ decay.

After production, these long-lived particles ($\Psi = N,a$) travel a distance $\ell_{\rm prod}$ before reaching a particle detector, where they can decay into the observable $e^+e^-$ pair final state that is targeted in this study.\footnote{In the case of HNL decays, the neutrino contributes to missing energy.} The probability for $\Psi$ to decay into $(\nu)e^+e^-$ inside this detector is given by
\begin{equation}
\mathcal{P} = e^{-\ell_{\rm prod}/L_\Psi} \left( 1 - e^{-\Delta \ell_{\rm det}/L_\Psi}\right) \mathcal{B}_{e^+e^-},
\label{eq:Prob}
\end{equation}
where $\Delta \ell_{\rm det}$ is the detector length along the particle trajectory, $L_\Psi$ is the boosted decay length of $\Psi$, and $\mathcal{B}_{e^+e^-} = \Gamma(\Psi\to (\nu)e^+e^-)/\Gamma_\Psi$ the branching ratio of $\Psi$ into the $e^+e^-$ signal channel with $\Gamma(\Psi \to (\nu)e^+e^-)$ the partial width of $\Psi$ to decay into the visible channel.
In the laboratory frame, this can be written as
\begin{equation}
L_\Psi = \gamma_\Psi \beta_\Psi c\,\tau_\Psi = c\,\tau_\Psi \frac{p_\Psi}{m_\Psi},
\end{equation}
where $p_\Psi$ is the magnitude of the momentum of the particle, $m_\Psi$ its mass, and $\tau_\Psi$ its lifetime in the rest frame.
In the limit where $\ell_{\rm prod},\Delta\ell_{\rm det} \ll L_\Psi$, which is true throughout the entire parameter space explored here, the probability of decay is then 
\begin{equation}
    \mathcal{P} \simeq c \frac{p_\Psi}{m_\Psi} \,\Delta \,\ell_{\rm det} \, \Gamma(\Psi \to (\nu) e^+e^-).
\end{equation}
Therefore, in this long-lived particle regime, only the decay width into the visible decay channel is important.

\begin{figure*}[t]

    \includegraphics[width=1\textwidth]{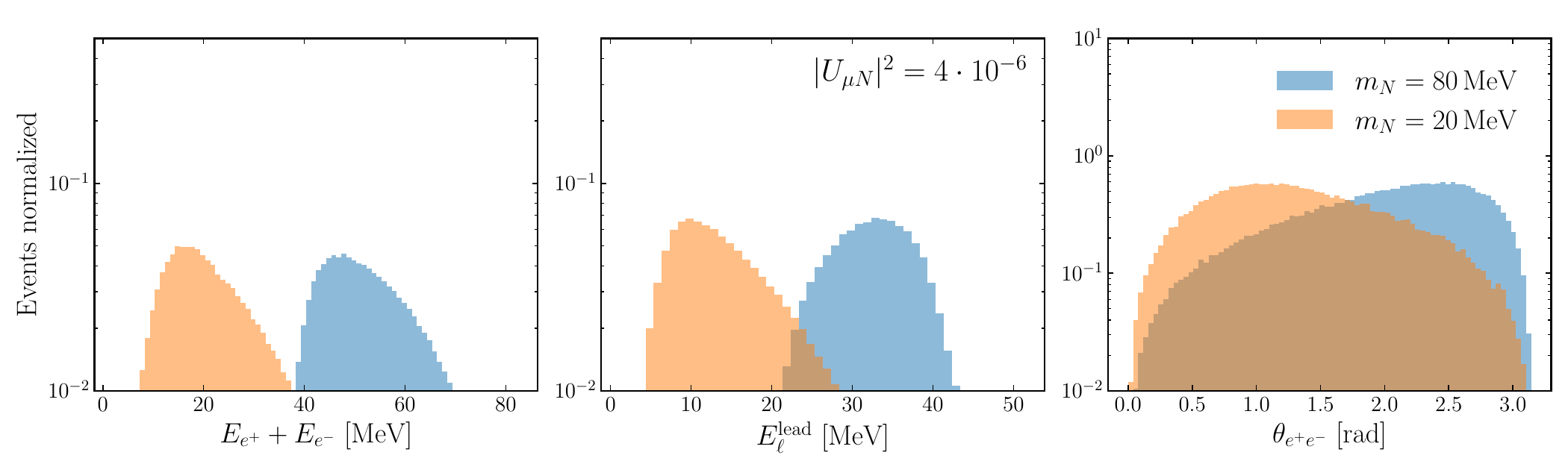}

    \includegraphics[width=1\textwidth]{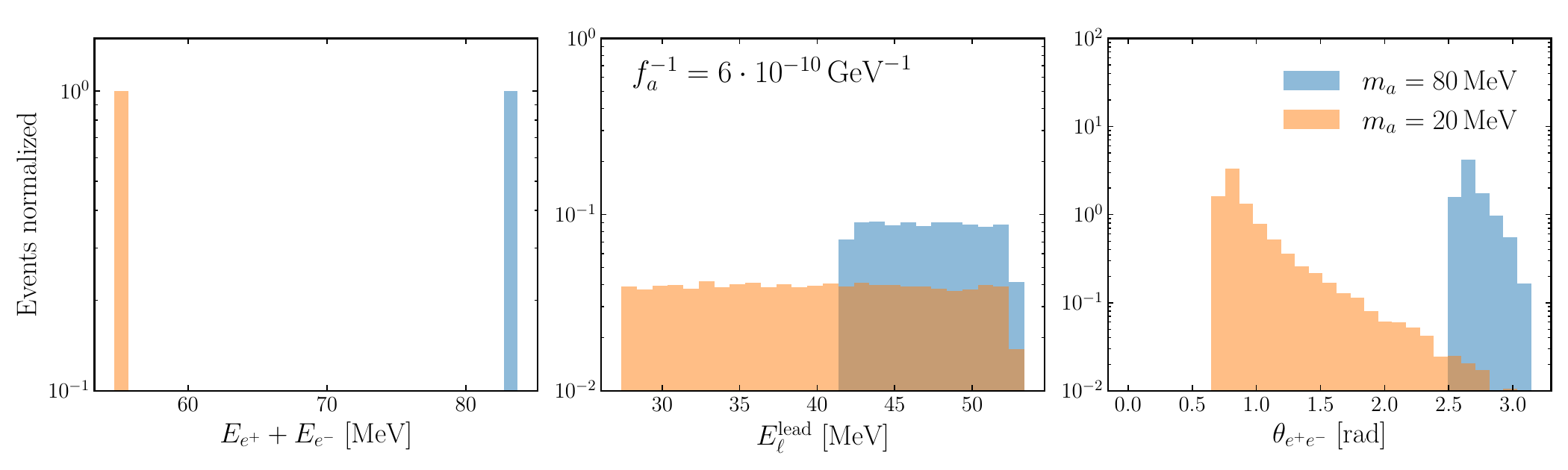}
    
    \caption{Normalized $e^{+}e^{-}$ kinematic distributions from heavy neutral lepton decays (upper row) and LFV ALPs (lower row). The figure shows truth-level distributions of the visible energy, $E_{e^+}+E_{e^-}$ (left), the leading lepton energy, $E_{\ell}^{\text{lead}}$ (middle), and the opening angle $\theta_{e^+e^-}$ (right), for two representative masses of 20 and 80 MeV.  Couplings are set to values close to the current experimental limits. \label{fig:signal_at_detector}}
\end{figure*}

In Figure~\ref{fig:signal_at_detector}, we present truth-level distributions of the total visible energy, the leading-lepton energy, and the opening angle for representative masses of 20 and 80 MeV. \footnote{For the case of HNLs, the event distributions have been obtained using the \texttt{MadGraph} HNL code developed in Ref.~\cite{Coloma:2020lgy}.}  
Final-state leptons have kinetic energies below but comparable in scale to the masses of the decaying LLPs.  
HNLs produce a broader and softer $e^+e^-$ summed energy spectrum due to their three-body kinematics, whereas ALPs exhibit sharper features reflecting their fixed boost from two-body production.  


\section{Experimental Parameters of a Potential HC$^2$ Campaign at SNS}
\label{sec:layout}

This study seeks to identify a viable experimental arrangement capable of delivering maximized sensitivity to ALP and HNL production in the SNS target.  
The phenomenology described in the previous Section thus calls for a large-volume detector in close proximity to the SNS target that is capable of reconstructing isolated $\mathcal{O}$(10-100)~MeV $e^+e^-$ pairs and distinguishing them from beam and cosmic backgrounds with high fidelity.  
Since the $\mu^+$ decay time is longer than the SNS beam spill, beam-related backgrounds can be reduced by focusing on detection of post-beam $e^+e^-$ signals.  
Steady-state cosmic backgrounds are expected to dominate the post-beam window, so a detector should feature technology that cleanly categorizes various signatures generated by on-surface, high-energy cosmic neutrons ($n$) and muons ($\mu^{\pm}$).
In the remainder of this Section, we will define the basic parameters of an SNS-based HC$^2$ campaign meeting these criteria.  

The SNS operates a 2~MW proton beam by directing 1.3~GeV protons onto a thick liquid-mercury target.  
These properties are updated with respect to prior BSM analyses~\cite{Hostert:2025ffy} due to the recent completion of ORNL's decade-long Proton Power Upgrade (PPU)~\cite{Plum:2018wms}.  
The beam has a repetition rate of 60~Hz, producing proton pulses with a duration of 0.4~$\mu$s. In addition to an intense neutron flux, the spallation process generates a large number of charged pions.   
A yield of approximately 0.09~$\pi^+$ per proton on target (POT) was estimated prior to the PPU beam upgrade~\cite{COHERENT:2021yvp}.  
For the purposes of this study, we assume a linear scale-up in pion production with beam energy~\cite{Grant:2015jva} post-upgrade, to 0.112~$\pi^+$ per POT.  
For final sensitivity estimates, we assume that the HC$^2$ detector observes 5$\times$10$^{23}$ POT of SNS beam operation, which corresponds to roughly 3~calendar years of HC$^2$ data-taking with 5000 operational beam hours per calendar year~\cite{Mammosser:2023dsn,Champion:2024vjj,Schedule}.

\begin{figure}[tbhp!]
    \includegraphics[trim = 0.0cm 0cm 0cm 0cm, clip=true, width=0.95\columnwidth]{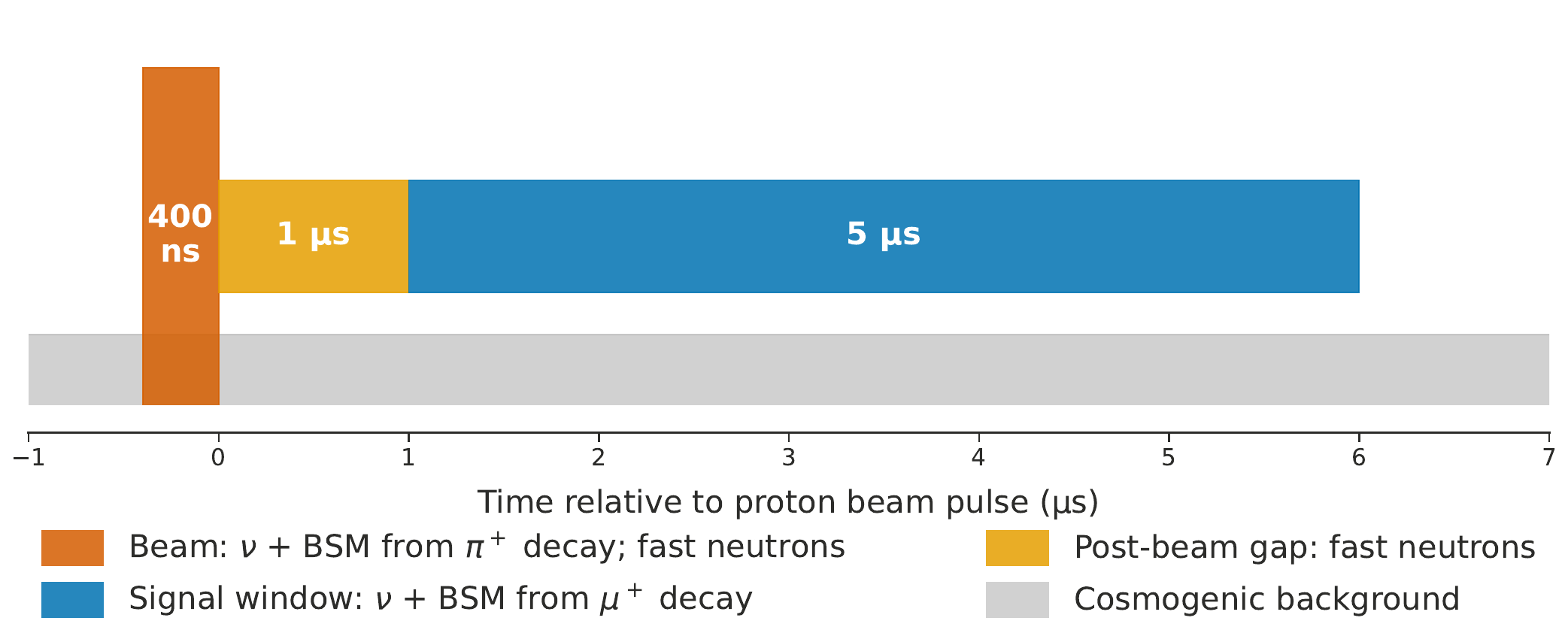}
    \caption{An illustration of the ordering of beam (orange) and LLP signal (blue) time windows.  The 5~$\mu s$ wide signal window targets detection of $e^+e^-$ produced after the beam window by decaying $\mu^+$ that were generated in $\pi^+$ decays during the beam window.  A steady state comic background (gray) is expected to be present during the signal window alongside a small contribution from $\mu^+$ decay neutrinos.} 
  \label{fig:time_structure}
\end{figure}

Pions produced by proton spallation in the target stop over $\mathcal{O}$(ns) timescales and decay at rest, subsequently producing a commensurate flux of $\mu^+$.  
We will assume that $\pi^+$ and $\mu^+$ decays occur centrally within the target bulk, with $\pi^+$ decays ($\mu^+$ production) occurring at a constant rate during the beam spill window.  
Most $\mu^+$ decays will occur after the end of the 0.4$~\mu$s SNS beam spill window due to the $2.2~\mu$s $\mu^{+}$ lifetime.  
To suppress prompt beam-related backgrounds while maximizing signal acceptance, we define a delayed signal analysis window from 1 to 6~$\mu$s after the proton pulse end.  
This time-ordering of beam and signal windows, illustrated in Figure~\ref{fig:time_structure}, indicates that steady-state comic backgrounds are expected to be present during the signal window alongside a small contribution from $\mu^+$ decay neutrinos; the latter sub-dominant contribution is briefly discussed in Section~\ref{sec:neutrino}.  
Over a nominal three-year data-taking period, this narrow timing window corresponds to a total integrated detector live time of approximately $1.6 \times 10^4$~seconds.  

While we will consider several potential LLP detector configurations for a HC$^2$ campaign at the SNS, the nominal case will roughly follow the example of the PROSPECT detector~\cite{prospect_nim}, which itself shares commonalities with the prior Bugey-3 reactor antineutrino detector~\cite{Abbes:1995nc}.  
We provide a brief overview of the nominal HC$^2$ configuration here and give further description of it and alternate cases in Section~\ref{sec:det}.  
A roughly 2$\times$2$\times$1.2~m$^3$ scintillator detector target will be assumed to host a 3.8~ton $^6$Li-loaded scintillator target subdivided into long, identical, and optically isolated segments.  
The scintillator's assumed timing characteristics and high light yield will allow determination of whether energy was deposited in a segment primarily by low-mass ($e^{\pm}$) or high-mass ($n$-$p$ scatters) charged particles~\cite{Aleksan:1988er,PROSPECT:2015rce,prospect_ls}.  
$^6$Li doping will allow for high-purity identification of thermal neutrons in the target~\cite{Aleksan:1988qh,PROSPECT:2018hzo}.  
Detector segmentation will afford rough but concrete information regarding the topology of an energy deposit, while also allowing clear identification of heavy and light charged particle sub-content within a larger event topology~\cite{Aleksan:1988er,prospect_prd}.  
Segmentation also unlocks capabilities for self-shielding within the relatively large detector volume. 
These design features enable a broad array of background rejection strategies beneficial for discriminating the LLP decay $e^+e^-$ pairs from omnipresent cosmogenic and radiogenic backgrounds at the SNS.  


\begin{figure}[tbhp!]
    \includegraphics[trim = 0.0cm 0cm 0cm 0cm, clip=true, width=0.85\columnwidth]{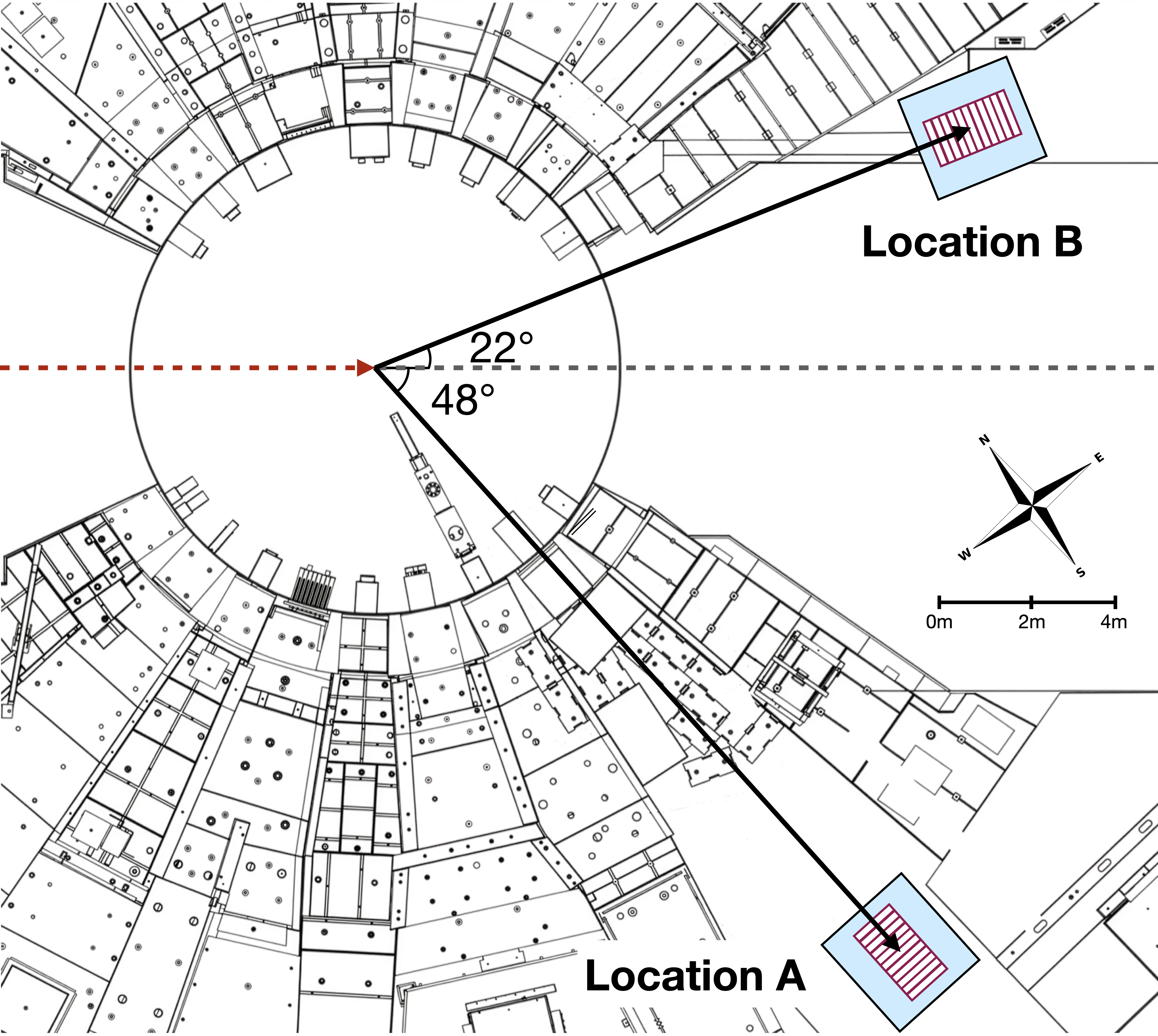}
    \caption{Overhead view of the SNS instrument floor showing two candidate deployment locations for the \HCdet detector.  
    Locations A (BL8, instrument floor) and B (basement, $\sim$6~m below) are each approximately 20~m from the SNS target. Black lines indicate the baselines to each location, forming angles of 48$^{\circ}$ (A) and 22$^{\circ}$ (B) relative to the beam direction (red dashed arrow). The detector is shown including external shielding, with the active volume in purple.} 
  \label{fig:position}
\end{figure}

The roughly 4~m$^3$ target size of this nominal HC$^2$ detector is likely too large to be deployed in the Neutrino Alley facility that hosts COHERENT's suite of detectors~\cite{Barbeau:2021exu,COHERENT:2021qbu}. 
Thus, other potential deployment locations must be considered.  
Two possible locations, both roughly 20~m removed from the SNS target, may be suitable for a HC$^2$ campaign, as depicted in Figure~\ref{fig:position}.
One location is in sector BL8 of the SNS instrument floor, a currently un-instrumented neutron beamline location 48$^{\circ}$ off-axis from the direction of the impinging SNS proton beamline.  
A temporary beam plug, plus a series of large temporary shielding blocks, provide a modest, non-hermetic shield for fast neutrons between this location and the SNS target.  
A second deployment spot is located in a modestly more on-axis location along the proton beamline (22$^{\circ}$)~\cite{sns_basement}.  
This location, in a basement utility area roughly 6~m below the elevation of the SNS target, offers many meters of rock and concrete fast beam neutron shielding, as well as modest ($<$few~m concrete) overburden from the building's structural and shielding materials.  
In the event that additional shielding is desired for reducing steady-state radiogenic and post-beam neutron capture and/or activation backgrounds, the BL8 location offers sufficient space and floor-loading capacity to host a detector inside of a large (10s of tons) passive shielding package.  
Space constraints are tighter for the basement location, so a large passive shielding package may or may not be feasible at this location.  

For LLP sensitivity studies, these two deployment locations are considered as interchangeable options, given their similar $\sim$20~m target-to-detector-center distance.  
In practice, differences in post-beam background and overburden between locations could be factors driving deployment location selection.
For LLP signal estimates, we assume that the nominal HC$^2$ detector is deployed such that the center of its active volume is at the same elevation as the SNS target, such that there is no overburden shielding the detector from cosmic ray fluxes.  
The detector's orientation is set such that its 2.0$\times$1.2~m$^2$ front face is normal to this center-to-center vector, with the axes of the 1.2~m-long horizontal active segments oriented perpendicular to this vector, as depicted in Figure~\ref{fig:position}.   

\section{Detectors, Datasets, and Simulations}
\label{sec:det}

In the nominal considered HC$^2$ campaign at SNS, a PROSPECT-like detector would aim to detect the scintillation light generated by final-state $e^+e^-$ pairs within the detector's active volume.  
In this Section, we summarize important detector design features of PROSPECT and other hypothetical HC$^2$ implementations and then introduce the real and simulated datasets to be used in this study.  

\subsection{The PROSPECT Detector}

PROSPECT consisted of an 11$\times$14 array of long liquid scintillator segments optically separated by mm-thick specular reflecting mirrors~\cite{prospect_grid}.  
Each segment consisted of 24~kg of $^6$Li-doped scintillator (LiLS~\cite{prospect_ls}) with a square cross-section 14.5~cm on a side and a length of 117~cm, with each long end bounded by an acrylic housing containing one 5" diameter photomultiplier (PMT).  
The full active detector array thus contained 3.72~t of scintillator within a total 3.8~m$^3$ volume.  
The active detector array and excess LiLS were contained inside a sealed acrylic vessel nested within a water-filled aluminum secondary containment tank.  
Beyond the 2.1$\times$2.0$\times$1.6~m$^3$ secondary tank, a shielding package consisting of nested volumes of boron-doped polyethylene (5-10~cm thickness), lead (2.5~cm thickness), polyethylene lumber (10~cm thickness) and a topping layer of water bricks (46~cm thickness) increased the general dimensions of the full detector and shielding package to 3.25$\times$2.95$\times$3.25~m$^3$.  
Specifics of the PROSPECT detector and shield can be found in Ref.~\cite{prospect_nim}.

Scintillation photons generated by particle interactions in a PROSPECT segment were transported efficiently to the segment-end PMTs, with total light collection varying from roughly 400 to 800 photoelectrons (PE) per MeV depending on position along a segment and on the date of data-taking.  
The fraction of light emission from short (16~ns) and long (38~ns) scintillation time constants of the LiLS varied substantially between hadronic ($n$-$p$ scatters, $\alpha$, $^3$H, and other heavy nuclei) and electromagnetic ($\gamma$, $e^-$ and $e^+$) energy depositions, causing these differing particle types to generate distinctive scintillation pulse shapes -- an attribute referred to as pulse shape discrimination (PSD).  
Following a detector trigger, full PMT waveforms were processed with 14-bit, 500~MHz CAEN v1725 waveform digitizers and zero-suppressed to enable trigger rates well in excess of 10$^4$ Hz~\cite{prospect_prd}.  

PMT HV and electronics gain settings in PROSPECT were tuned to balance strong PSD performance at the lowest accessible energies ($\sim$100~keV) with linear energy response beyond the highest energies relevant to reactor antineutrino detection ($\sim$10~MeV).  
As a result of this optimization, recorded PMT waveforms became saturated when the total energy deposited in a PROSPECT segment exceeded a few dozen MeV.  
This means that, for example, downward-going cosmic $\mu^{\pm}$ passing through PROSPECT, as well as hard proton scatters of $>$100~MeV cosmic neutrons, were likely to have generated clipped or distorted waveforms in one or more electronics channels, which biased or limited downstream reconstruction capabilities.  
In addition, some PROSPECT PMTs experienced HV-related failures resulting from LiLS leakage into  mineral oil filled PMT housings~\cite{Andriamirado:2021qjc}.  
Over the course of PROSPECT data-taking, this resulted in a gradually increasing number of segments featuring one or zero functional PMTs, which reduced or eliminated affected segments' reconstruction capabilities.  
For these reasons, the validated performance of PROSPECT described below represents a conservative estimate of the ultimate performance of an HC$^2$ campaign at the SNS.  

\subsection{HC$^2$ Detector Configurations}

The design parameters of a specific considered HC$^2$ implementation exist in a large multi-dimensional parameter space of scintillator technology options with varied target size, sub-element ($i.e.$ segment) size, photo-detector coverage, light collection, PSD capability, neutron capture target, dynamic range, and more.  
To ground our study in a practical and previously demonstrated portion of this technology phase space, we adopt nominal HC$^2$ detector parameters similar to those of PROSPECT. 
Similar detector parameter choices also allow outcomes from PROSPECT cosmic simulation-data comparisons to be  confidently extrapolated to the HC$^2$ context.  
Thus, this HC$^2$ detector features the same number of segments and similar segment cross-sections, PMT instrumentation (all fully functional), scintillator and optical response properties, and readout electronics and triggering system attributes.  
To optimize for a HC$^2$ data-taking campaign focused on LLP decay products at the SNS, the dynamic range is assumed to be shifted from PROSPECT's $\mathcal{O}$(20~keV-20~MeV) range to the $\mathcal{O}$(200~keV-200~MeV) range.  

It is also interesting to explore the performance of other practically accessible scintillator technology phase space regions.  
For this reason, we also consider HC$^2$ detector implementations featuring changes to target liquid and segmentation properties: 
    \begin{itemize}
        \item \textbf{Un-doped LS:} Doping LS with $^6$Li may improve event selection by reducing the target's $n$-capture time constant and by ensuring that $n$-capture clusters generate easily-recognizable, high-PSD clusters.  Given the time, cost, and potential risk associated with $^6$Li doping~\cite{prospect_ls}, it is worth considering how a HC$^2$ detector can perform with commercially-available, undoped LS.  
        \item \textbf{Non-PSD LS:} It is worth considering the use of low-cost LS options that do not feature any PSD capability. This option allows us to study how a reduction in the ability to tag $n$-$p$ scatters impacts downstream physics.   
        \item \textbf{Monolithic target:} Optical or physical target segmentation can improve event selection by enhancing knowledge about the topological characteristics of physics signatures in HC$^2$.  However, it also necessitates production of a dedicated segmentation subsystem~\cite{prospect_grid}, increases the amount of material in contact with the LS, and limits the placement and configuration of light sensors.  Given these practical challenges, as well as the existence of numerous past and potential future single-volume hydrogenated detector targets at spallation neutron sources~\cite{LSND:1996jxj,COHERENT:2021xhx,JSNS2:2021hyk,Anderson:2022lbb,COHERENT:2026ewu}, we also consider a HC$^2$ case where we use an event selection matched to the attributes of a monolithic detector target.  
    \end{itemize}
If a HC$^2$ detector is designed to facilitate target re-configuration, emptying, and refilling, it could potentially host multiple experimental arrangements over its lifetime.   
We end by noting that we have not exhaustively explored background rejection performance over the full potential technology phase space; some attributes that would be interesting to explore in future studies are pure water or water-based LS~\cite{Ford:2022wla,Xiang:2024jfp} target liquid, hybrid Cerenkov/scintillation photo-sensing capabilities~\cite{Caravaca:2016ryf}, and more finely segmented target implementations~\cite{Haghighat:2018mve,Sutanto:2021xpo,LiquidO:2024piw,MobileAntineutrinoDemonstratorProject:2025sna}.  

%

\subsection{Simulation and Dataset Descriptions}

For this study, we will consider a subset of PROSPECT data collected when the HFIR reactor was shut down for refueling.  
This data, acquired during reactor-off periods in 2018 between April 20 to June 5, contains around $2.9 \times 10^9$ total triggered events.  
With an on-surface location at HFIR featuring $<$1 meters water-equivalent of overburden, PROSPECT datasets during reactor-off periods consist entirely of radiogenic and cosmogenic backgrounds.  
With 377 hours of data-taking, this dataset's live time is roughly 85 times larger than the $1.6 \times 10^4$~seconds of data expected to be collected at SNS during the 5~$\mu$s post-beam $\mu^+$ decay window.  
Of PROSPECT's 154 segments, 20 (2) were viewed by one (two) PMT(s) experiencing HV issues at some point during the data period; of these, 10 (2) segments were located along the active detector boundary.  
Data from partially- or non-functioning segments will not be considered in subsequent analysis for this paper.  

To facilitate the design of an optimized shielding package and to better understand its reactor-off backgrounds, the PROSPECT collaboration developed and validated Monte Carlo (MC) simulations for on-surface cosmic $n$ and $\mu^{\pm}$, called SG4~\cite{PROSPECT:2015iqr}.  
Based on the Geant-4 particle transport code, SG4 performs a two-step procedure that enables simulation of rare cosmic ray interaction processes in a computationally efficient manner.  
In the first simulation step, cosmic ray $\mu^{\pm}$ or $n$ with parameterized angular distributions and energy spectra matching state-of-the-art surface-based measurements are generated on a continuous surface above a simulated experiment geometry consisting of the detector, its shielding package, and some surrounding building features.  
Cosmic primaries and secondaries are then propagated through the experiment geometry while recording only the properties of particles crossing the boundary of a detector's active target region.  
A second round of MC simulation then faithfully regenerates boundary-crossing particle combinations, propagates them into the active detector region, and applies realistic detector response features.  
As described in Ref.~\cite{prospect_prd}, SG4 (referred to in Ref.~\cite{prospect_prd} as `PG4,' an earlier iteration of the same simulation package) has simulated response effects that are tuned using calibration constant databases from real PROSPECT datasets.  
The output of SG4-simulated events matches that of PROSPECT's raw data files, such that SG4 outputs can be processed identically to data in PROSPECT's reconstruction and analysis framework.  

For this study, we simulate 195 (11) total hours of cosmic $n$ ($\mu^{\pm}$) fluxes in PROSPECT.  
For HC$^2$ cosmic ray simulations, we produced few-hour overburden-free cosmic ray MC datasets for two different configurations described in the previous section: 
    \begin{itemize}
        \item \textbf{Case A, Low-Gain LiLS:} Identical detector response to a fully-functional PROSPECT detector, but with waveform saturation effects turned off.  This case approximates the expected response of a low-gain run of a HC$^2$ detector at the SNS.  
        \item \textbf{Case B, Low-Gain LS:} Identical to Case A, but with $^6$Li removed from the LS material properties file.  This case represents a HC$^2$ run with commercial undoped but PSD-capable LS.  
    \end{itemize}
As will be described in Section~\ref{sec:select}, other variations in HC$^2$ detector design will be studied through adjustments in analysis cuts, as opposed to adjustments in the simulated geometry.  

To determine expected signal rates for the various LLP scenarios, we also generated a set of HC$^2$ MC simulations containing $e^+e^-$ pairs with kinematic properties as described in Section~\ref{sec:flux}.  
These simulations used dedicated two-body and three-body decay generators that incorporated the relevant phenomenology and relativistic kinematics of HNLs and ALPs that would be produced at the SNS.  
Separate MC simulations were performed for a few discrete values of LLP mass to assess variations in realized signal selection efficiency across the relevant phase space.  
To reflect the HC$^2$ placement in Figure~\ref{fig:position}, $e^+e^-$ kinematics are assigned orientations matching the assumption that BSM particles enter the detector with a velocity roughly normal to the front face of the detector.  
Decay vertices are generated uniformly throughout the active HC$^2$ detector, ignoring the $\sim$30\% variation between front and back detector faces due to 1/$\ell_{\rm prod}^2$ flux fall-off, as well as coupling/mass dependent variations in the decay length $L_\Psi$.

\section{BSM Particle Signal Selection}
\label{sec:select}

In this section, we define the physics reconstruction and reconstructed quantities of interest for this analysis, and then give a description of the analysis cuts we apply to select particle interactions in an HC$^2$ detector matching the expected appearance of final-state $e^+e^-$ pairs from BSM particle decays.  

\subsection{Event Reconstruction}

Reconstruction of acquired PMT waveforms in HC$^2$ data proceeds by first determining the attributes of energy depositions in individual segments, and then by determining broader facets of a particle's entire interaction by combining information from different segments. Detailed descriptions of this event reconstruction paradigm are given in Ref.~\cite{prospect_prd}.  

Information from recorded waveforms for both PMTs within a segment are processed to make a \emph{pulse} data object, which is assigned reconstructed properties describing the particle interactions in that segment.  
By comparing time and charge ratios between time-aligned PMT waveforms, an event's position can be reconstructed along the segment length $z_{rec,p}$ with resolution comparable to that provided in orthogonal dimensions by the target's optical segmentation.  
Summed, time-aligned waveform shapes are used to determine a pulse shape discrimination metric (PSD$_p$) that reflects the fractional area ratio of the waveform's tail.  
Summed amplitudes plus knowledge of $z_{rec,p}$ is used to reconstruct the energy $E_{rec,p}$ of the pulse.  
Saturation of a pulse's PMT waveforms can generate a downward bias in $E_{rec,p}$ values and an upward bias in PSD$_p$, while $z_{rec,p}$ values, primarily driven by waveform arrival time differences, are much less affected.  
Since scintillation light collection is more even between PMTs for energy depositions in the center of a segment, this analysis sometimes features fiducial cuts in $z_{rec,p}$ to ameliorate potential waveform saturation effects.  

If pulses from different segments have leading waveform edges aligned within $\pm$20~ns, they are grouped to form \emph{cluster} objects.  
Clusters retain the reconstructed information of its individual pulses while being assigned higher-level physics metrics.  
In this analysis, we make use of a cluster's reconstructed energy ($E_{rec,c}$, the sum of all $E_{rec,p}$ values in a cluster), its reconstructed segment ($S_{rec,c}$, the segment with the highest $E_{rec,p}$), its aggregate PSD value ($PSD_{c}$, the $E_{rec,p}$-weighted sum of all $PSD_{p}$ values), and its multiplicity (the total number of pulses it contains).  


\begin{figure}[tbhp!]
   \centering
   \includegraphics[width=0.85\columnwidth]{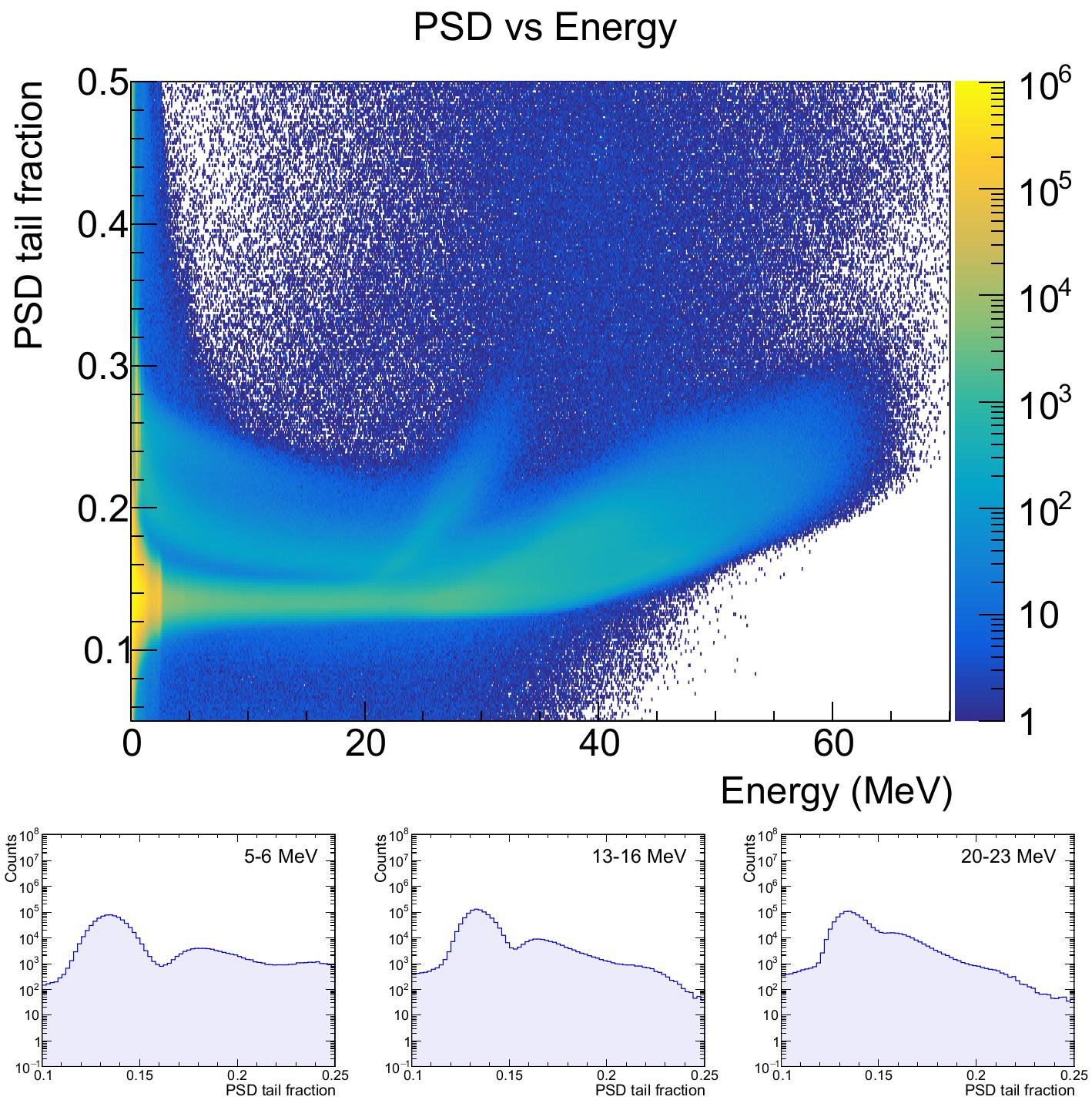}
    
    
    
    
    \caption{Top: reconstructed $E_{\text{rec},p}$ versus $\text{PSD}_p$ values for 
    all $z_{\text{rec},p}$-fiducialized single-pulse clusters (cluster multiplicity 
    of 1) with $|z_{\text{rec},p}|< 30$~cm in PROSPECT reactor-off data. Bottom: one-dimensional PSD 
    distributions across different vertical $E_{\text{rec},p}$ slices.}
    \label{fig:psd_vs_energy}
\end{figure}

To demonstrate some properties of reconstructed cluster and pulse physics metrics, we show in Figure~\ref{fig:psd_vs_energy} reconstructed $E_{rec,p}$ versus $PSD_{p}$ values for all single-pulse clusters (cluster multiplicity of 1) with $|z_{rec,p}|<30$~cm (where $z=0$ corresponds to the segment center) from the PROSPECT reactor-off dataset used in this study.
The clear PSD-based separation of electromagnetic (bottom band) and hadronic (top bands) energy deposits is seen to degrade as waveform saturation effects begin to set in for some pulses in PROSPECT with $E_{rec,p}>$20~MeV.  

\subsection{Signal Selection Criteria}

As illustrated in Figure~\ref{fig:signal_at_detector}, the final-state decay $e^+$ and $e^-$ are expected to have kinetic energies in the vicinity of 5 to 60~MeV, which, in hydrocarbon LS with an expected density close to 1 g/cm$^3$, corresponds to ranges of order 2 to 30 cm.  
Given the 14.5~cm segment width of the nominal considered HC$^2$ case, one should expect that most $e^+e^-$ primaries will experience substantial collisional energy losses (direct ionization) within 1 or a few segments.  
Meanwhile, the $\sim$100~MeV critical energy of hydrocarbon materials indicates that radiative losses (bremsstrahlung $\gamma$-ray emission), while relatively commonplace, will usually comprise a minority of the total energy of the primaries.  
In addition, the final-state $e^+$ will also generate annihilation $\gamma$-rays.  
Considering the expected $\mathcal{O}$(10~cm) attenuation length of MeV-scale $\gamma$-rays in LS, one should expect that secondary bremsstrahlung and annihilation $\gamma$-rays will commonly generate energy deposits in one or more segments beyond those hosting the BSM particle decay.  


Backgrounds are expected to be dominated by primary interactions from cosmic ray neutrons and $\mu^{\pm}$ in or near the active detector volume.
Detector fiducialization can be expected to reduce backgrounds from directly incident $\mu^{\pm}$ and, to a lesser extent, backgrounds from directly incident neutrons and $\gamma$-rays, as well as from spallation products generated by cosmic rays interacting in material surrounding the detector.
In addition, primary or muon-induced spallation neutrons within the detector volume will predominantly generate hadronic energy deposits, either via energetic $n$-$p$ scattering or thermal $n$ capture on $^6$Li.

\begin{figure}[htb!pb]
\centering

\includegraphics[trim=2.0cm 95.0cm 2.0cm 10.0cm, clip=true, width=0.95\columnwidth]{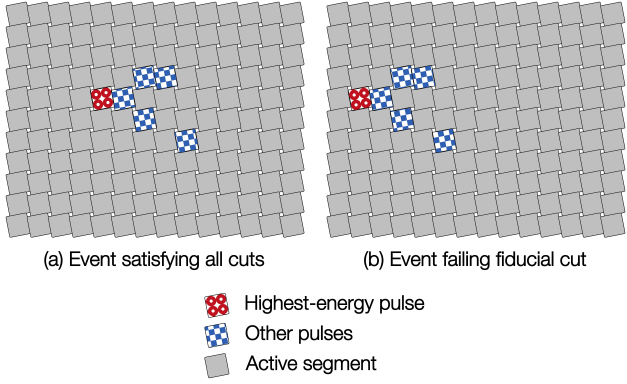}

\caption{Examples illustrating topology and fiducialization event selection requirements. Colored cells indicate detector segments containing reconstructed pulses, with the highest-energy pulse shown in red. Panel (a) shows an event satisfying all selection criteria. Panel (b) shows an event rejected by the fiducial requirement. 
}

\label{fig:example_fig}
\end{figure}

These expected background features and electromagnetic physics considerations, combined with rough optimization studies on subsets of PROSPECT data and HC$^2$ $e^+e^-$ MC, direct us to apply the following signal cluster selection cuts, which are represented graphically in Figure~\ref{fig:example_fig}: 
\begin{itemize}
    \item{It must have a multiplicity between 2 and 10 pulses.}    
    \item{All $PSD_{p}$ values must be within 3$\sigma$ of the $\gamma$-like PSD band.}  
    \item{To fiducialize, the most-energetic pulse cannot have $S_{rec,p}$ in the two outer-most top (side) segment rows (columns) or in the three bottom-most rows.  This pulse must also have $|z_{rec,p}|<$30~cm. All other pulses are allowed to occur in any segment, as long as $|z_{rec,p}|<$50~cm.}  
\end{itemize}

Some cosmogenically-produced clusters matching the selection requirements may be produced alongside other time-coincident signatures.  
Thus, any candidate signal cluster is vetoed if: 

\begin{figure*}[htb!pb]
\centering

\includegraphics[trim=0.2cm 0.1cm 0.0cm 0.2cm, clip=true, width=0.33\textwidth]{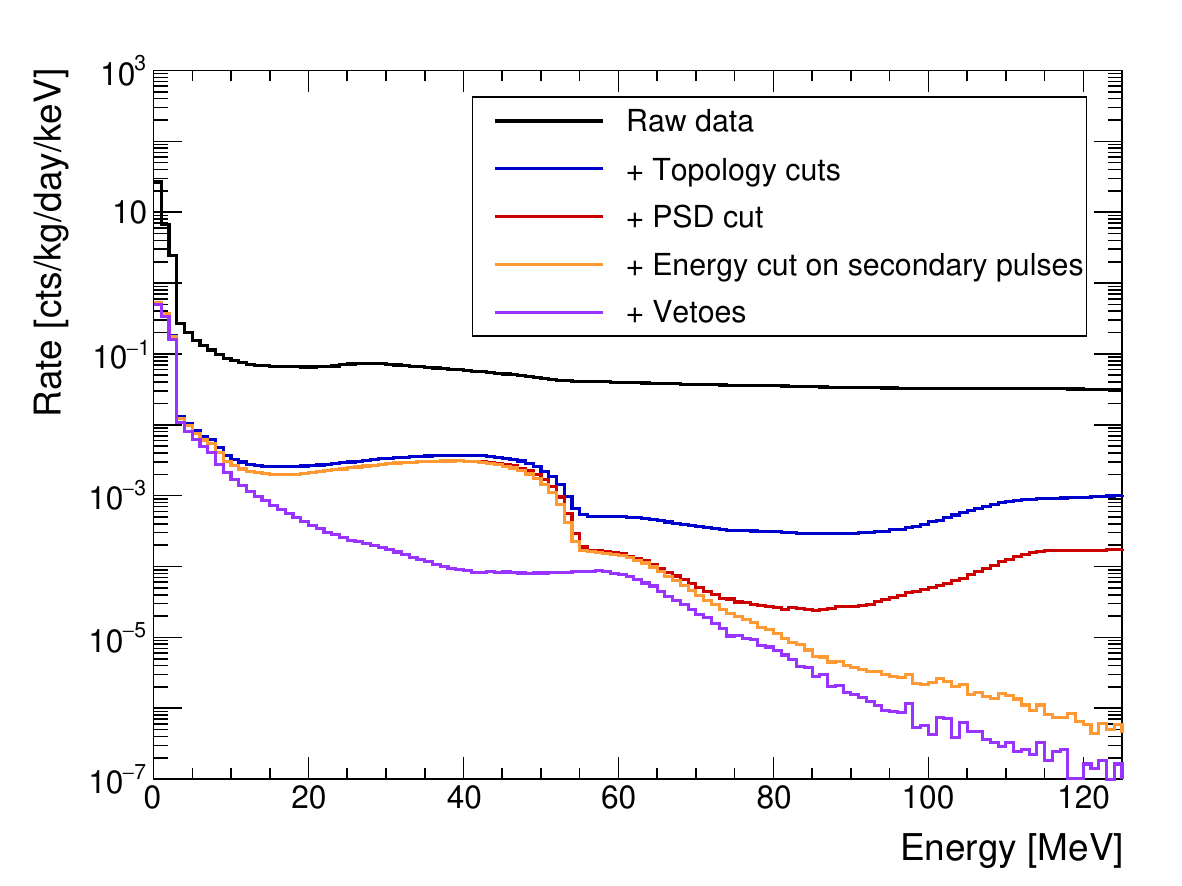}
\hfill
\includegraphics[trim=0.2cm 0.1cm 0.0cm 0.2cm, clip=true, width=0.28\textwidth]{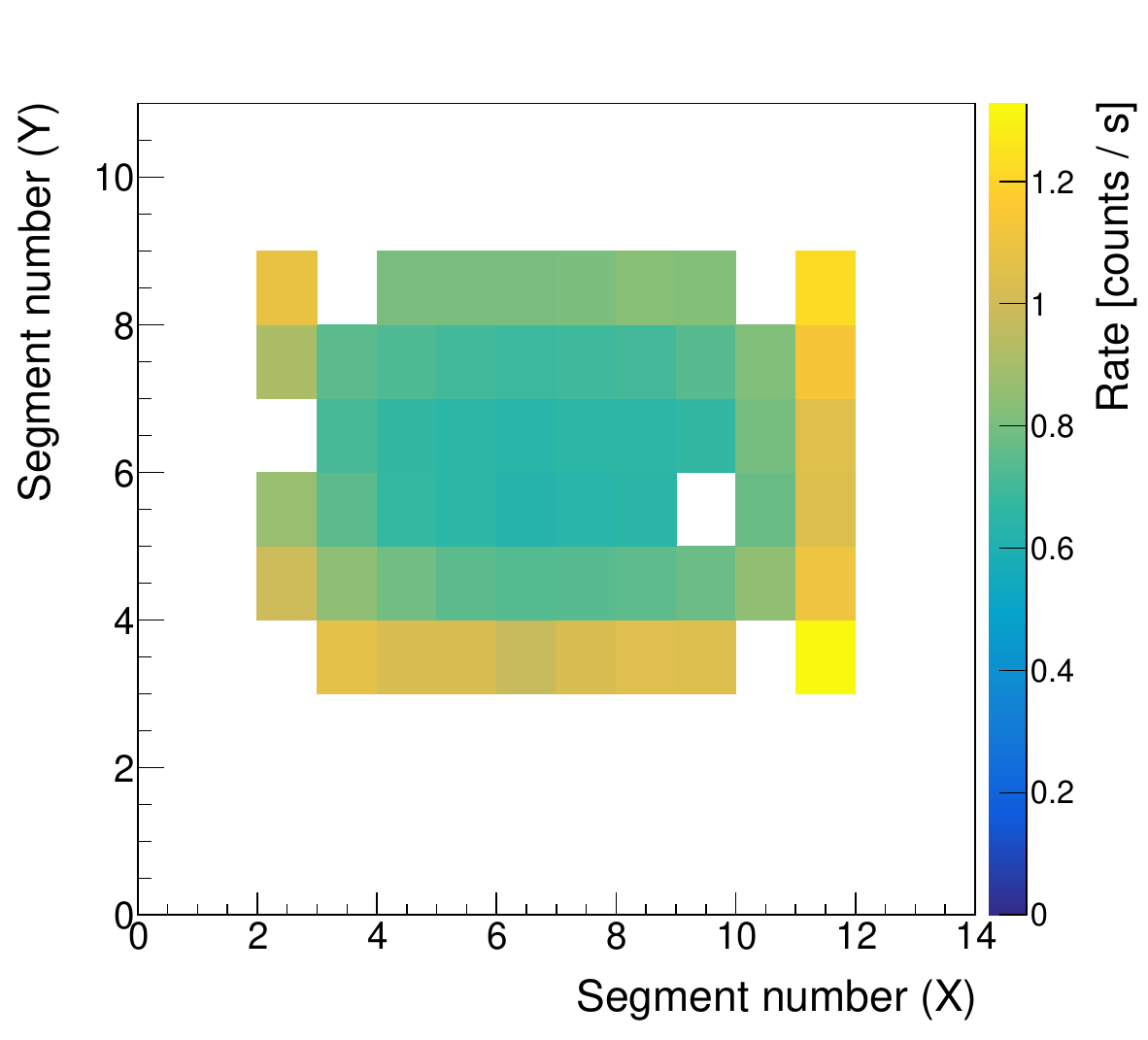}
\hfill
\includegraphics[trim=0.2cm 0.1cm 0.0cm 0.2cm, clip=true, width=0.34\textwidth]{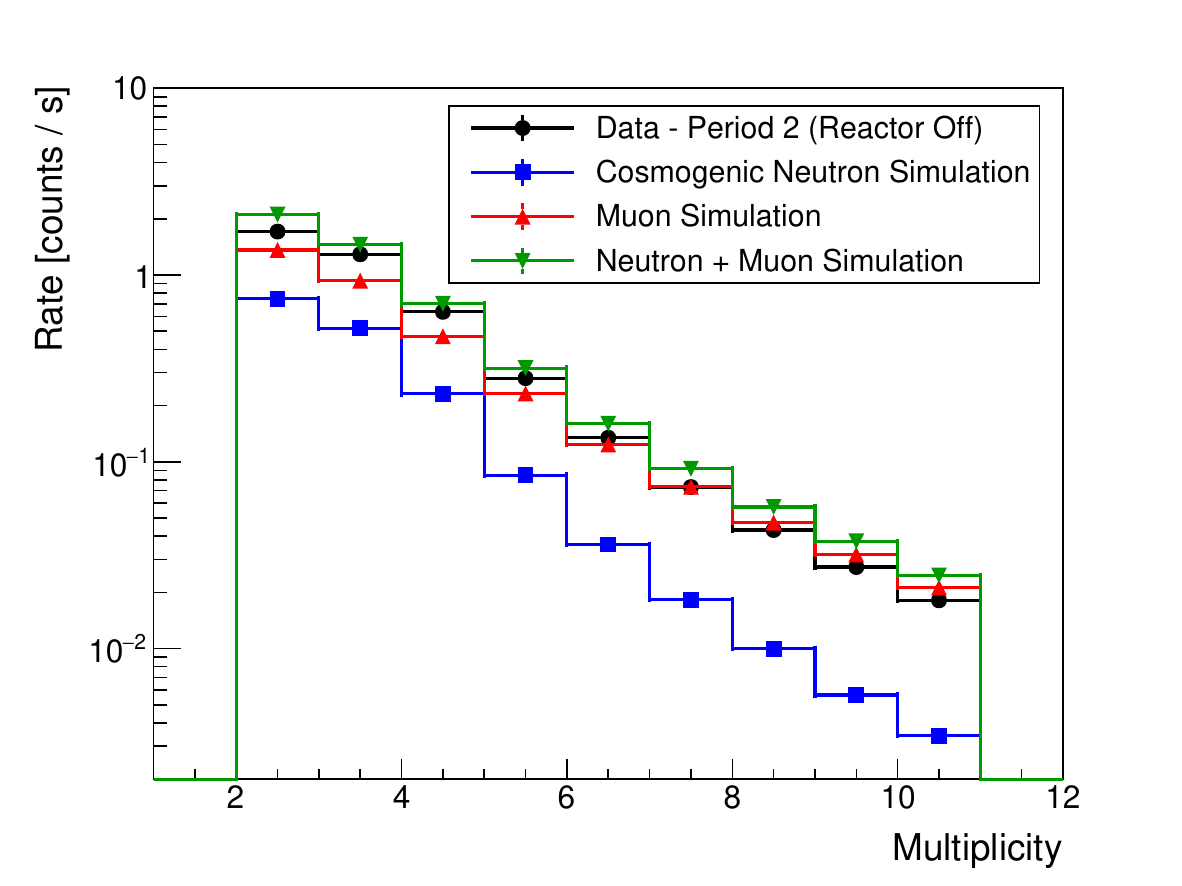}


\caption{Attributes of selected $e^+e^-$ candidate signal clusters in PROSPECT data and cosmic MC simulations after application of analysis cuts. Left: reconstructed cluster energy $E_{rec,c}$ for PROSPECT data at various stages of selection cut application. Middle: segment hosting the highest-energy pulse in a cluster ($S_{rec,c}$) for PROSPECT data. Right: reconstructed pulse multiplicities of candidate clusters for PROSPECT data and cosmic ray MC simulations. To facilitate comparison with the cosmogenic MC samples, only PROSPECT data clusters with reconstructed energies $E_{rec,c} \geq 2.5$~MeV are shown in the right panel, reducing the impact of low-energy background populations not modeled in the cosmogenic MC simulations. Error bars represent statistical uncertainties.}

\label{fig:P1_data}
\end{figure*}

\begin{itemize}
    \item It occurs within 100~$\mu$s after a cluster with $E_{rec,c}>$15~MeV.  This cut is aimed at vetoing $\mu^{\pm}$ decay and spallation products. The observed muon-like rate of roughly 500~Hz in PROSPECT~\cite{prospect_prd} indicates an associated dead time of $\sim$5\% in a similarly-sized, on-surface HC$^2$ implementation.  
    \item It occurs within 600~$\mu$s before a cluster matching the signature of thermal neutron capture on $^6$Li (a single-pulse cluster with the correct PSD-energy combination~\cite{prospect_prd}).  PROSPECT's on-surface $^6$Li capture rates ($\sim$ 2~Hz) indicate $<$0.1\% dead time for this cut in HC$^2$.    
     \item It occurs within 600~$\mu$s before a cluster matching the expected attributes of $n$ capture on $^1$H, $^{12}$C, $^{35}$Cl, or other near-target materials (any cluster with $E_{rec,c}>2~MeV$ and all $|z_{rec,p}|<$50~cm).  PROSPECT reactor-off cluster rates of $\sim$50~Hz above 2~MeV indicate a dead time of 3\% from this veto cut.  
    \item It occurs within 0.8~$\mu$s of any other cluster.  This cut is aimed at rejecting potentially mis-reconstructed clusters appearing in the same readout window as another cluster. A reactor-off trigger rate of 1600~Hz in PROSPECT illustrates that $<$0.1\% associated dead time can be achieved with this kind of requirement.  
\end{itemize}

As will be described in the following sections, these baseline signal selection and background reduction criteria can be adjusted according to the specifics of the HC$^2$ implementation under consideration.  
Given the general similarity between ALP and HNL decay products, we apply identical cuts for both LLP searches.  
It is likely that dedicated selections optimized for specific LLP types and masses could further improve projected HC$^2$ sensitivities.  
Furthermore, given the multi-dimensional nature of pulse and cluster reconstructed quantities, optimized signal-background categorization would likely benefit from application of boosted decision tree or deep learning AI/ML methods comparable to those previously applied to low-level data objects in other neutrino experiment contexts~\cite{Aurisano:2016jvx,DUNE:2020gpm,MicroBooNE:2021tya,MicroBooNE:2021zai}.  

\section{PROSPECT Results and Cosmic Data-Simulation Comparisons}
\label{sec:results1}


We begin presentation of event selection results by demonstrating outcomes on data from a realized implementation of a tons-scale HC$^2$ detector: PROSPECT.  
PROSPECT results demonstrate the utility of the selection and cosmic background modeling tools, building confidence in the robustness of subsequent HC$^2$ background estimates and LLP search sensitivities in the SNS context.  
For PROSPECT, the baseline selection criteria above must be adapted to account for waveform saturation effects, which, as demonstrated in Figure~\ref{fig:psd_vs_energy}, begin to manifest themselves in PROSPECT's fiducial region above roughly 20~MeV $E_{rec,p}$.  
To address this, the per-pulse $PSD_p$ requirement is not applied to the maximum-energy pulse if its $E_{rec,p}$ is greater than 20~MeV.  
In addition, apart from the maximum $E_{rec,p}$, all other $E_{rec,p}$ must be $<$20~MeV.  

Figure~\ref{fig:P1_data} and Table~\ref{tab:dru} show the results of application of all selection cuts to PROSPECT data.  
In the 377~h dataset, a total of 304310 candidate events are observed from 20-100 MeV $E_{rec,c}$, with rates of 1.54$\times$10$^{-4}$ and 2.95$\times$10$^{-5}$ events/kg/keV/day (DRU) in the 20-50~MeV and 50-100 MeV regimes, respectively.  
Over the 20-100~MeV range, applied cuts reduce total signal rates by a factor of $\sim$10$^{3}$.  
For reference, JSNS$^2$ reports detecting  2.20$\times$10$^{-4}$ $e^+$-like events with between 20-60~MeV per 8\,$\mu$s post-spill window in its 12.3~m$^3$ ($\sim$10~t) fiducial volume, or 5$\times$10$^{-3}$ DRU~\cite{JSNS2:2023hxl}.

\begin{table*}[htpb!]
  \caption{Rate of detection of various cosmic background cluster categories in PROSPECT data and cosmic MC simulations. Rates are given in DRU (events/kg/keV/day) prior to any selection cuts, after application of signal selection criteria, and after application of time coincident veto cuts.}
    \begin{tabular}{l|cc|cc|cc}
    \hline \hline
          & \multicolumn{2}{c}{Raw Rate (DRU)} & \multicolumn{2}{|c|}{Event Rate (DRU)} & \multicolumn{2}{c}{Post-Veto Rate (DRU)} \\
           & 20-50~MeV & 50-100~MeV  & 20-50~MeV & 50-100~MeV & 20-50~MeV & 50-100~MeV  \\
          \hline
    PROSPECT, Data      & 6.21$\times 10^{-2}$ & 3.67$\times 10^{-2}$ & 2.62$\times 10^{-3}$ & 1.17$\times 10^{-4}$ & 1.54$\times 10^{-4}$ & 2.95$\times 10^{-5}$ \\ \hline
    PROSPECT, $n$ MC    & 4.05$\times 10^{-3}$ & 1.56$\times 10^{-3}$ & 1.07$\times 10^{-4}$ & 5.12$\times 10^{-5}$ & 3.54$\times 10^{-5}$ & 2.29$\times 10^{-5}$ \\
    PROSPECT, $\mu$ MC  & 5.29$\times 10^{-2}$ & 3.20$\times 10^{-2}$ & 2.64$\times 10^{-3}$ & 7.94$\times 10^{-5}$ & 1.27$\times 10^{-4}$ & 8.32$\times 10^{-6}$ \\\hline
    PROSPECT, Total MC  & 5.69$\times 10^{-2}$ & 3.35$\times 10^{-2}$ & 2.74$\times 10^{-3}$ & 1.31$\times 10^{-4}$ & 1.63$\times 10^{-4}$ & 3.12$\times 10^{-5}$ \\
    \hline \hline
    \end{tabular}%
  \label{tab:dru}%
\end{table*}%

\begin{figure}[htb!pb]
\includegraphics[trim=0.5cm 0.1cm 1.0cm 0.1cm, clip=true, width=0.85\columnwidth]{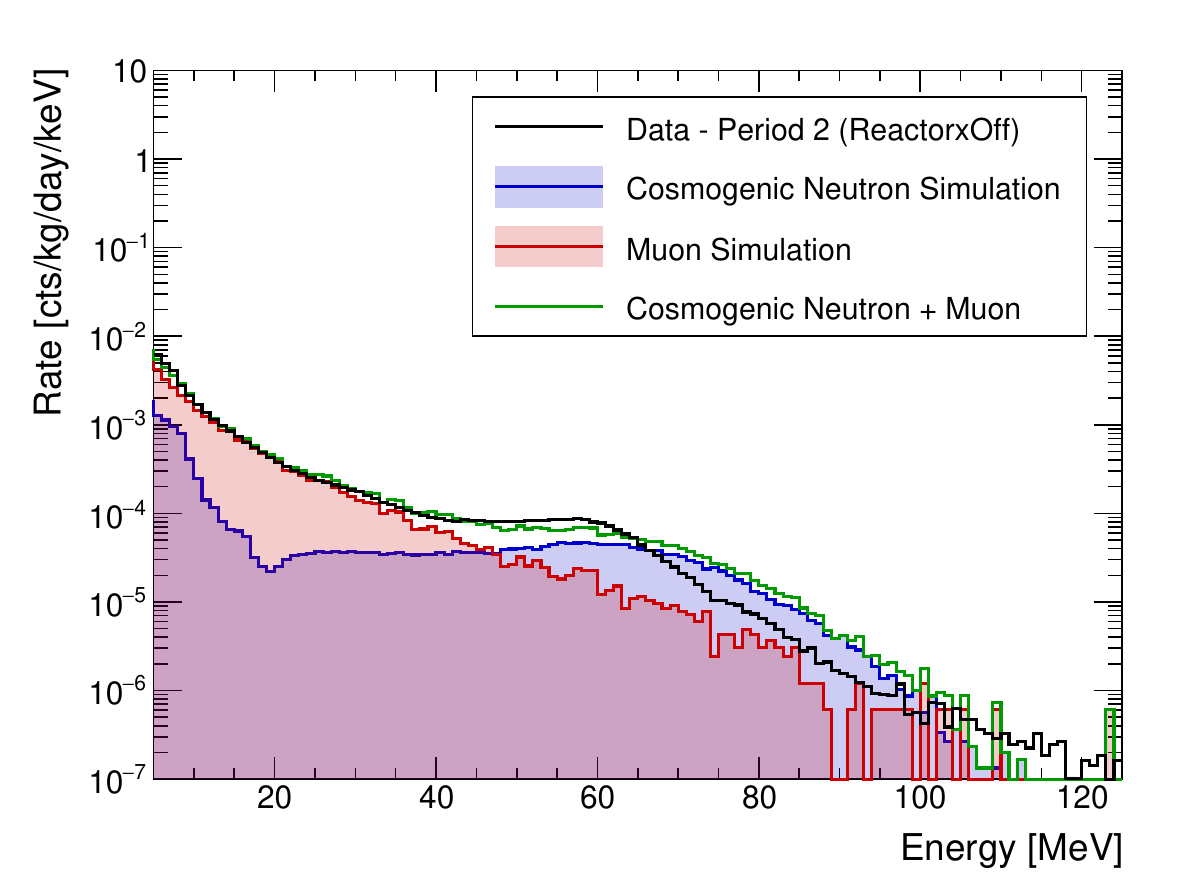}
\caption{Reconstructed cluster energy $E_{rec,c}$ for PROSPECT data and cosmic MC simulation ($\mu^{\pm}$ and $n$) normalized to DRU (cts/kg/day/keV) after application of all signal selection and veto cuts.}
\label{fig:P1_dataMC}
\end{figure}


As shown in the spectrum of cluster energies $E_{rec,c}$ in the left panel from Figure~\ref{fig:P1_data}, the combination of all signal selection cuts (the orange curve in Figure~\ref{fig:P1_data}) isolates a large sample of cosmic $\mu^{\pm}$ decay electrons, as evidenced by the energy spectrum cutoff at $\sim$55~MeV. This sample is entirely reduced after applying all veto cuts, leaving a fairly flat $E_{rec,c}$ spectrum with a broad shoulder peaking around 60~MeV caused by the saturation effect piling events with a broad true deposited energy range into a comparatively narrower range of $E_{rec,p}$.

Figure~\ref{fig:P1_data} also shows the cluster multiplicity and reconstructed segment $S_{rec,c}$ for selected events.  
These plots indicate a preference for smaller cluster size within the signal sample with a fairly even distribution of signal counts across the functioning fiducial segments.  

The same set of selection and veto cuts were also applied to simulated PROSPECT cosmic $\mu^{\pm}$ and $n$ datasets.  
Livetime-normalized results from these selections are also shown in Figures~\ref{fig:P1_data} and~\ref{fig:P1_dataMC} and Table~\ref{tab:dru}.  
We find that $e^+e^-$-like signal cluster rates in PROSPECT are primarily induced by cosmic $\mu^{\pm}$ below 50~MeV and by cosmic $n$ above 50~MeV.  
Total signal cluster rates between 20 and 100~MeV are consistent between data and simulation within a factor of 10\%, with better agreement ($\sim5\%$ differences) found in the $\mu^{\pm}$-dominated regime.  
The largest data-simulation offsets are found near the 60~MeV shoulder appearing in PROSPECT data, indicating imperfect modeling of waveform saturation effects in SG4.  
Given the difficulty of understanding true primary cosmic neutron fluxes in complex on-surface environments and in simulating neutron transport through these environments~\cite{1369506,BECCHETTI20151,PROSPECT:2015iqr,MicroBooNE:2024prh}, this reasonable level of data-simulation agreement is a strong indication that cosmic $n$ and $\mu^{\pm}$ interactions are the dominant contributor to high-energy $e^+e^-$-like rates in the PROSPECT reactor-off dataset.  
We also conclude that SG4 cosmogenic $n$ and $\mu^{\pm}$ MC simulations can be used to reliably estimate cosmogenic backgrounds for a LLP search in a HC$^2$ campaign at SNS.  
We end by noting that precise background estimates for a real HC$^2$ campaign would arise from off-beam data-taking, rather than dead-reckoned MC simulation results; thus, the residual data-MC offsets in Figure~\ref{fig:P1_dataMC} should not be viewed as a direct future hurdle to the performance of precise BSM searches at the SNS.  

\section{HC$^2$ Cosmic Simulation Results}
\label{sec:results2}


\begin{figure}[hbt!]
\centering

\includegraphics[trim=0.9cm 0.0cm 0.0cm 0.0cm, clip=true, width=0.85\columnwidth]{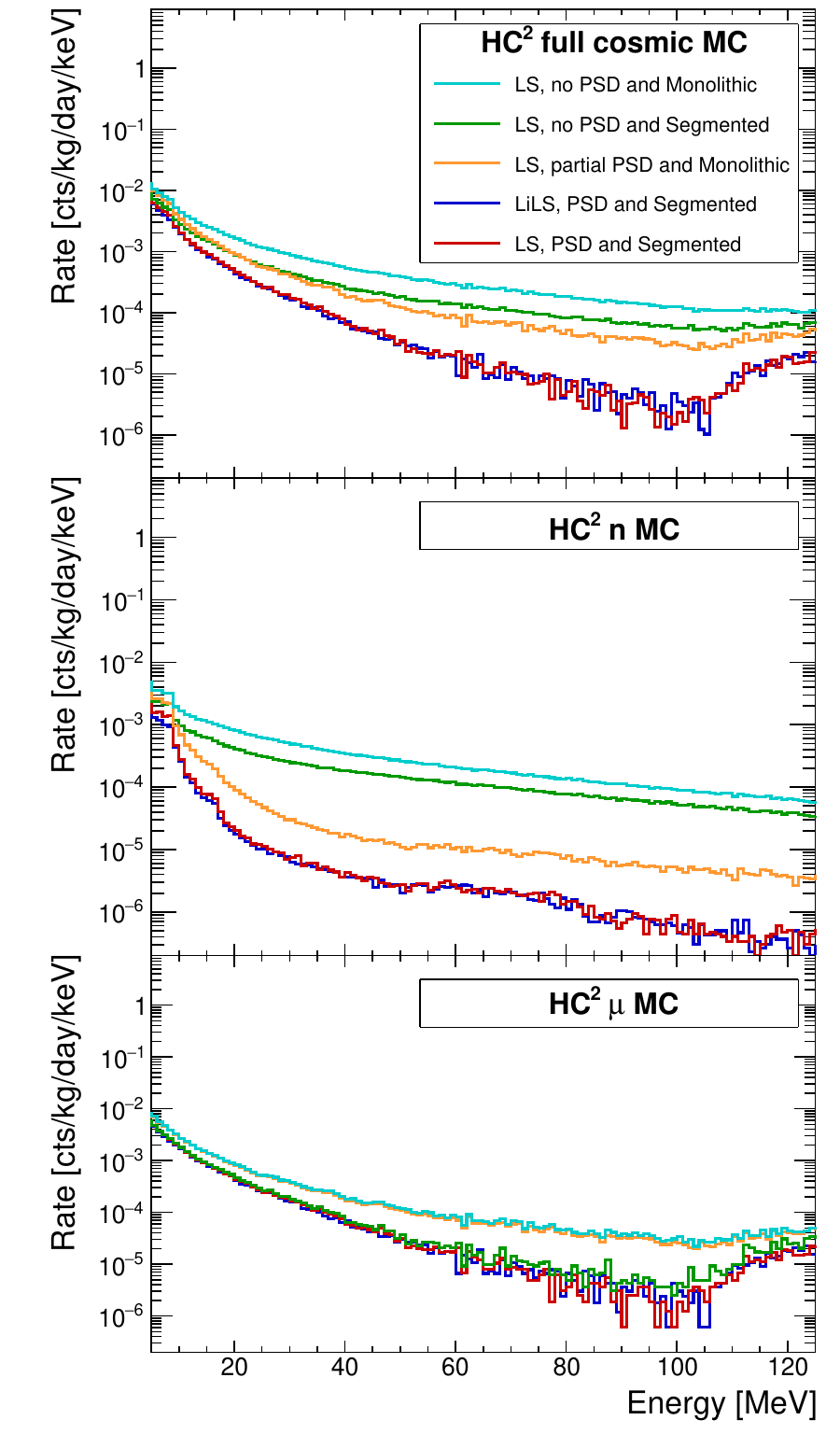}
\caption{Reconstructed cluster energy $E_{rec,c}$ for all the different HC$^2$ detector variations. Top:  Cosmic $n$ plus $\mu^{\pm}$ MC simulation. Middle: Only cosmic $n$ simulation. Bottom: Only cosmic $\mu^{\pm}$ MC simulation.  }

\label{fig:HC_MC}
\end{figure}

Having benchmarked the general accuracy of cosmic MC simulations with PROSPECT datasets, we turn to assessing expected cosmic backgrounds for an on-surface HC$^2$ data-taking campaign at the SNS.  
As mentioned in Section~\ref{sec:det}, the nominal HC$^2$ detector can be assumed to operate in a lower-gain mode free of waveform saturation effects at high $E_{rec,p}$, with this difference reflected in cosmic MC simulations.  
We also studied how the variations in HC$^2$ detector design described in Section~\ref{sec:det} might affect signal cluster counts and $E_{rec,c}$ spectra.  
To do so, in addition to the previously-described variations to HC$^2$ simulation, we consider adjustments to signal selection cuts that approximate other changes to HC$^2$ target fill liquid and segmentation properties: 

    \begin{itemize}
        \item \textbf{Case I, PSD:} PSD requirements apply to all pulses.  
        \item \textbf{Case II, No PSD:} All $PSD_{p}$ requirements are removed.  
        \item \textbf{Case III, PSD, No Segmentation:} Cluster multiplicity and maximum/secondary $E_{rec,p}$ requirements are removed, and $PSD_{p}$ requirements are substituted for a single 3$\sigma$ requirement on $PSD_{c}$.  The fiducialization requirement on the maximum $S_{rec,p}$ is kept, since this capability is likely to remain in some form in a monolithic detector target.  
        \item \textbf{Case IV, No PSD, No Segmentation:} Same as Case III, but with the $PSD_{c}$ requirement removed. 
    \end{itemize}

\begin{table*}[htpb!]
\centering
  \caption{Rate of detection of various cosmic background cluster categories in a HC$^2$ detector MC simulations for two different target material cases. Rates are given in DRU (events/kg/keV/day) and counts for a roughly 3~calendar years  after application of signal selection criteria, and time coincident veto cuts.}
  \begin{tabular}{l|cc|cc}
    \hline \hline
          & \multicolumn{4}{c}{Post-Veto Rate} \\
          & \multicolumn{2}{c|}{DRU (events/kg/keV/day)} & \multicolumn{2}{c}{Counts} \\
          & 20--50~MeV & 50--100~MeV & 20--50~MeV & 50--100~MeV \\
    \hline
    HC$^2$, $n$ PSD-LiLS (A-I)            & $6.34\times10^{-6}$ & $1.60\times10^{-6}$ & 133   & 56   \\
    HC$^2$, $\mu$ PSD-LiLS (A-I)          & $1.38\times10^{-4}$ & $9.26\times10^{-6}$ & 2894  & 324  \\\hline
    HC$^2$, $n$ PSD-LS (B-I)              & $6.89\times10^{-6}$ & $1.68\times10^{-6}$ & 142   & 58   \\
    HC$^2$, $\mu$ PSD-LS (B-I)            & $1.46\times10^{-4}$ & $8.68\times10^{-6}$ & 2999  & 297  \\\hline
    HC$^2$, $n$ LS (B-II)                 & $2.30\times10^{-4}$ & $8.95\times10^{-5}$ & 4724  & 3064 \\
    HC$^2$, $\mu$ LS (B-II)               & $1.58\times10^{-4}$ & $1.22\times10^{-5}$ & 3245  & 418  \\\hline
    HC$^2$, $n$ Monolithic PSD LS (B-III)     & $2.95\times10^{-5}$ & $8.21\times10^{-6}$ & 606   & 281  \\
    HC$^2$, $\mu$ Monolithic PSD LS (B-III)   & $3.12\times10^{-4}$ & $5.21\times10^{-5}$ & 6408  & 1783 \\ \hline
    HC$^2$, $n$ Monolithic LS (B-IV)    & $4.48\times10^{-4}$ & $1.58\times10^{-4}$ & 9201 & 5409  \\
    HC$^2$, $\mu$ Monolithic LS (B-IV)  & $3.30\times10^{-4}$ & $5.90\times10^{-5}$ & 6778 & 2020  \\
    \hline \hline
  \end{tabular}%
  \label{tab:dru_HC}%
\end{table*}


Results of cosmic MC simulations for the various HC$^2$ detector configurations are shown in Figure~\ref{fig:HC_MC} and Table~\ref{tab:dru_HC}.  
For a nominal case of a HC$^2$ detector with segmentation and fill liquid design features closest to that of PROSPECT (Case A-I in Table~\ref{tab:dru_HC}), we see signal rates of 1.44$\times10^{-4}$ (1.09$\times10^{-5}$) DRU in the 20-50 (50-100)~MeV range.  
These DRU correspond to a total of 3027 (380) total signal-like background counts in these same expected energy ranges for the full 3~year SNS run period of the nominal HC$^2$ campaign.  
These rates are lower than simulated rates in PROSPECT.  
Lower rates at high energies in HC$^2$ -- particularly the $>$10$\times$ reduction in primary neutron backgrounds -- highlight the benefits of improved HC$^2$ PSD performance due to the removal of saturation effects.  

Interestingly, as shown in Figure~\ref{fig:HC_MC}, cosmic $\mu^{\pm}$-induced backgrounds dominate $n$-induced backgrounds at all energy ranges below 100~MeV in HC$^2$: the former background component exceeds the latter by a factor of $>$20 at the lowest considered energies to a factor of 3 at around 100~MeV. 
This result highlights the value of including an external cosmic $\mu^{\pm}$ veto sub-system in a future HC$^2$ deployment.  
While this design feature was not included in PROSPECT for its reactor \nuebar campaigns at HFIR, it is included in many larger current- and next-generation COHERENT detectors~\cite{COHERENT:2021xhx,COHERENT:2024axu,COHERENT:2026ewu}.  

If a muon veto subsystem reduces $\mu^{\pm}$-induced backgrounds to sub-dominant levels, it seems likely that a full HC$^2$ campaign could achieve cosmic background levels on the order of hundreds of total counts.
Given the order-of-magnitude fall-off in primary cosmic neutron flux from 0 to~5~meters waters equivalent overburden~\cite{Gaisser:2016uoy}, the considered SNS basement deployment location may offer the promise of total cosmic backgrounds in the $<$100~count regime.  

Figure~\ref{fig:HC_MC} and Table~\ref{tab:dru_HC} indicate which HC$^2$ design attributes are crucial for cosmic background reduction in the $>$10~MeV energy regime.  
Background rates are similar between $^6$Li-doped and un-doped PSD LS cases.  
Thus, while the more costly and R\&D-heavy $^6$Li doping design element was crucial for PROSPECT's HFIR-based on-surface reactor \nuebar use case, it does not offer an obvious benefit in our SNS BSM context.  
This statement is contingent on the details of the $n$-capture vetoes described in Section~\ref{sec:select}: if post-beam activation or $n$-capture $\gamma$-ray backgrounds at SNS prove to be much higher than PROSPECT's steady-state reactor-off $\gamma$-ray backgrounds (roughly 50~Hz above 2~MeV $E_{rec,c}$), dead time in an un-doped HC$^2$ detector may rise to an unacceptable level.  

In contrast to the doping design element, removal of either the PSD element or the segmentation element of HC$^2$ design greatly increases both $\mu^{\pm}$-induced and $n$-induced backgrounds.  
Removal of PSD capabilities, in particular, results in a $>$30$\times$ increase in $n$-induced background, causing this category to dominate over $\mu^{\pm}$-induced backgrounds at all energy ranges.  
Removal of detector segmentation increases both $\mu^{\pm}$-induced and $n$-induced categories by factors ranging from 2 to 6.  
If both of these features are removed at the same time, backgrounds are again dominated by $n$-induced events, with a $\sim$100$\times$ increase in their rates with respect to the nominal detector design.  

\begin{figure}[hbt!]
\centering

\includegraphics[trim=0.9cm 0.0cm 0.0cm 0.0cm, clip=true, width=0.85\columnwidth]{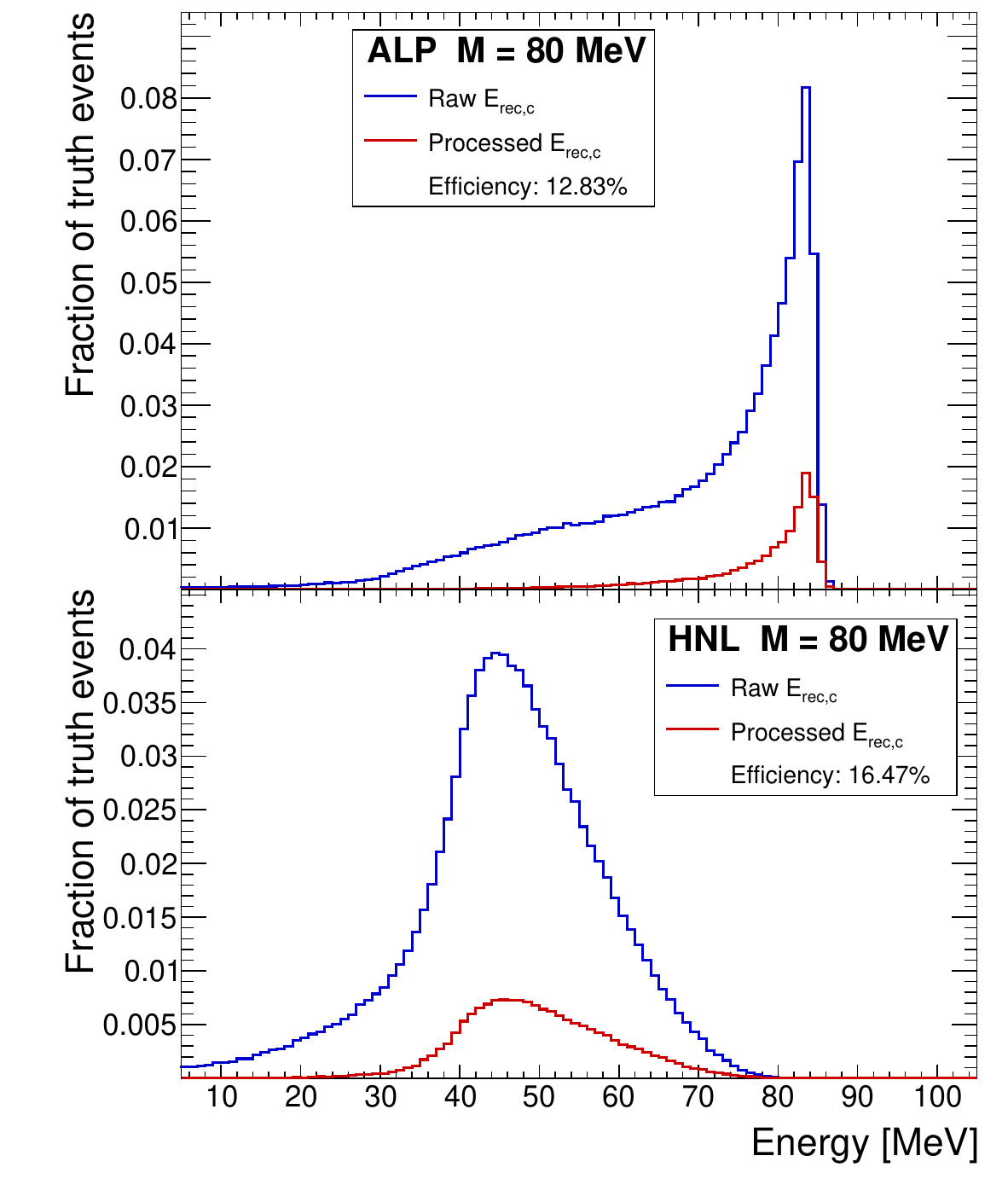}

\caption{Reconstructed energy spectrum before (blue) and after (red) our signal selection criteria for ALP (top) and HNL (bottom) for a mass of 80~MeV, truth spectrum is shown in Figure \ref{fig:signal_at_detector}. }

\label{fig:HC_signalMC}
\end{figure}

\begin{figure*}[t]
    \centering

    \includegraphics[width=0.45\textwidth]{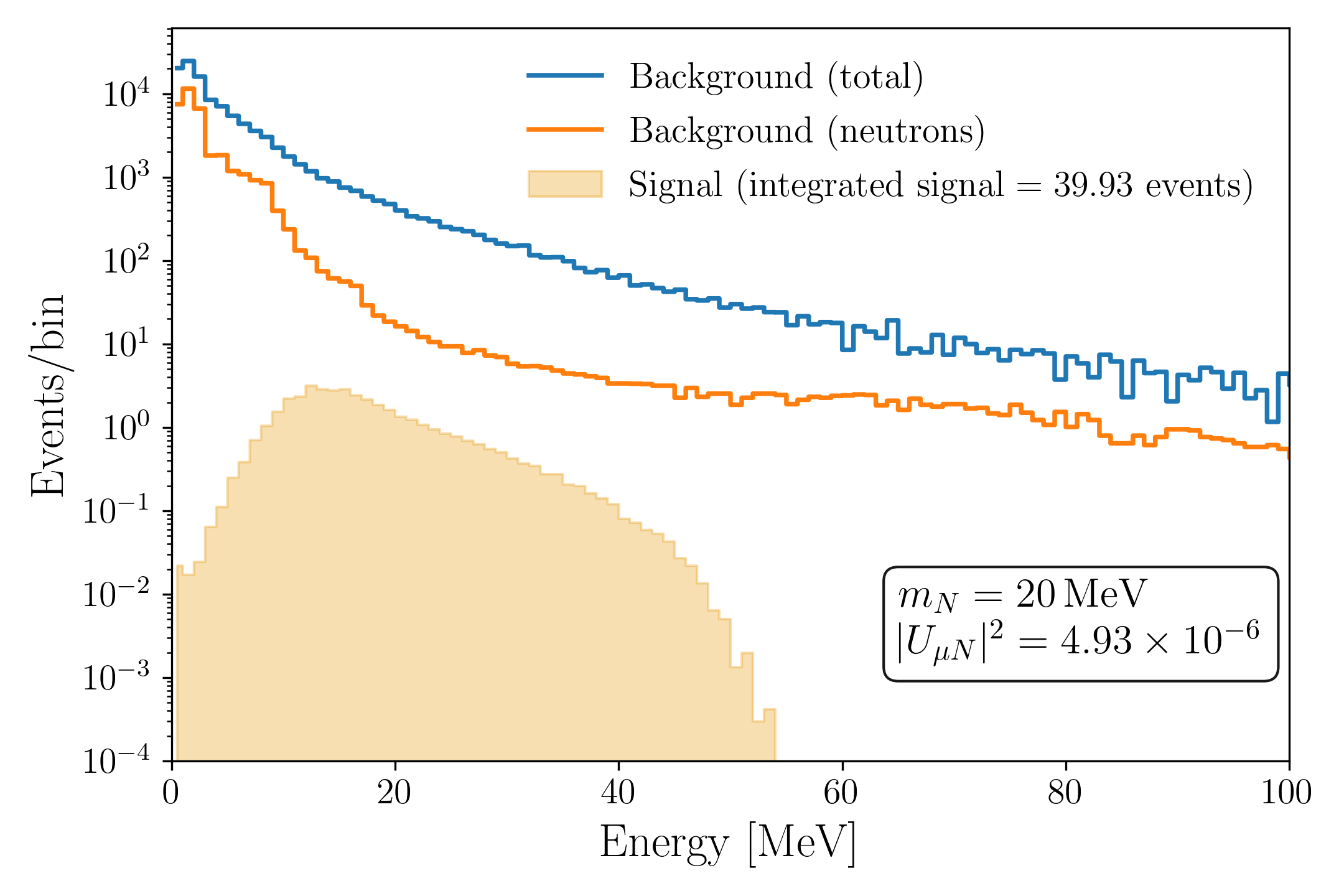}
    \includegraphics[width=0.45\textwidth]{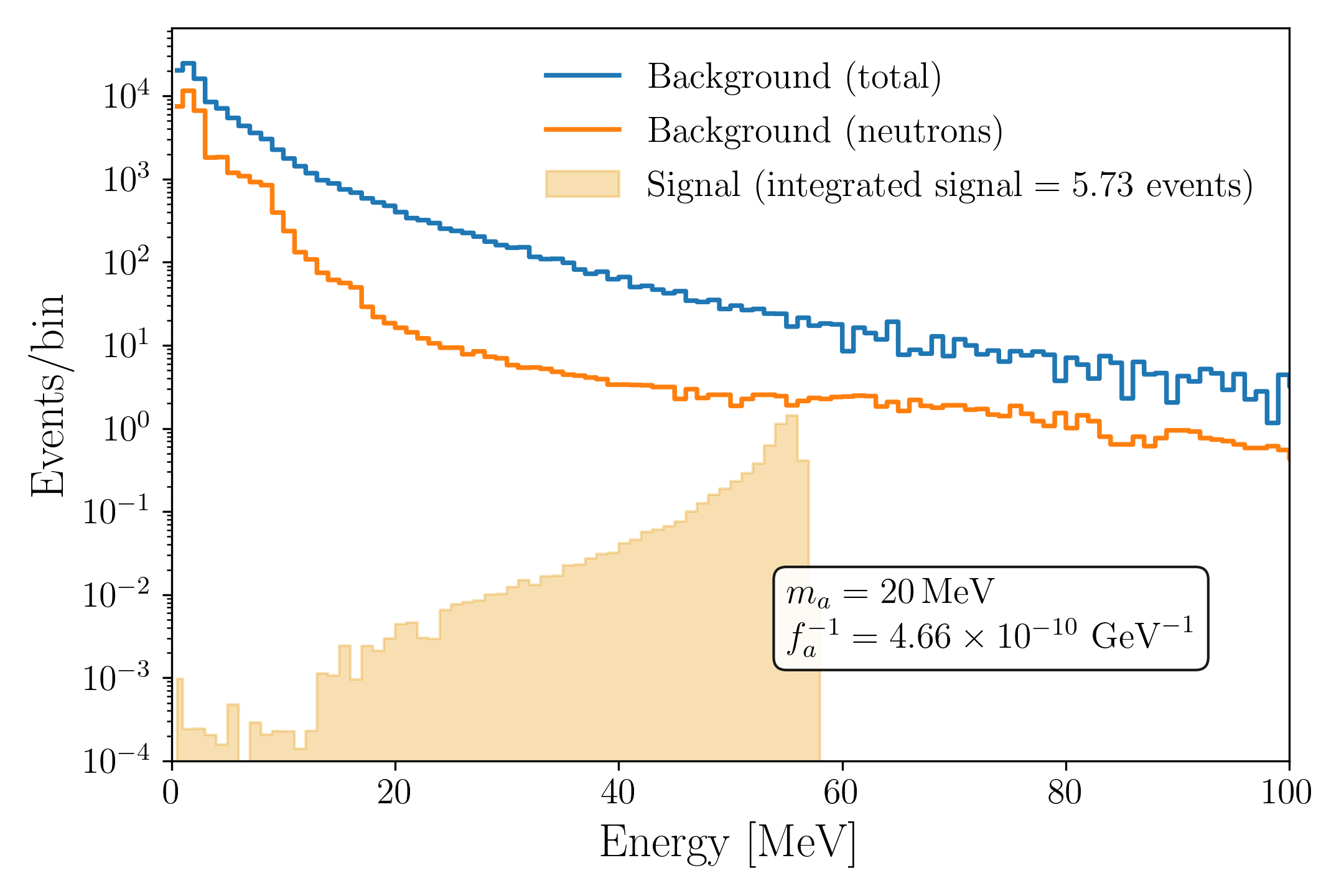}

    \vspace{0.3cm}

    \includegraphics[width=0.45\textwidth]{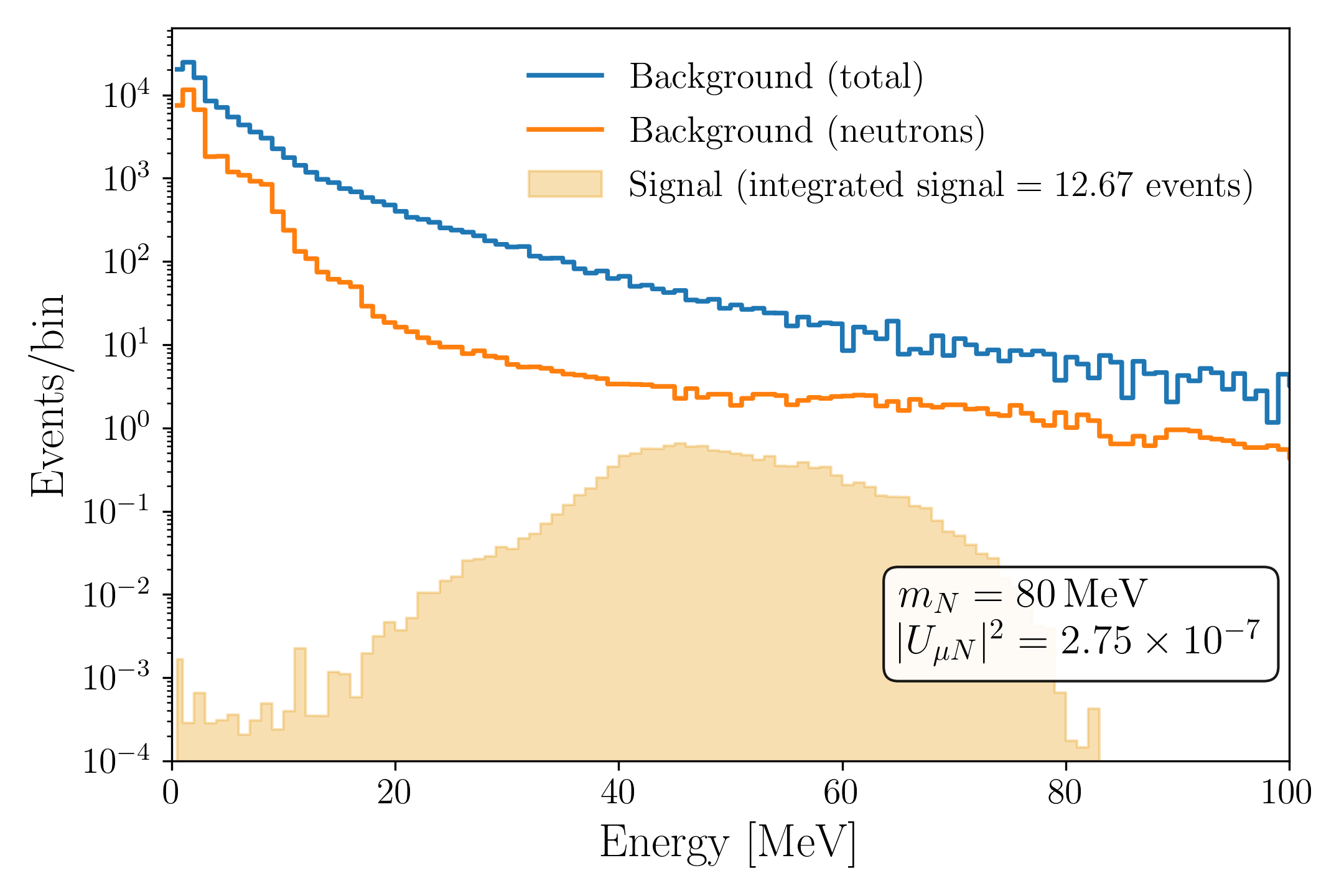}
    \includegraphics[width=0.45\textwidth]{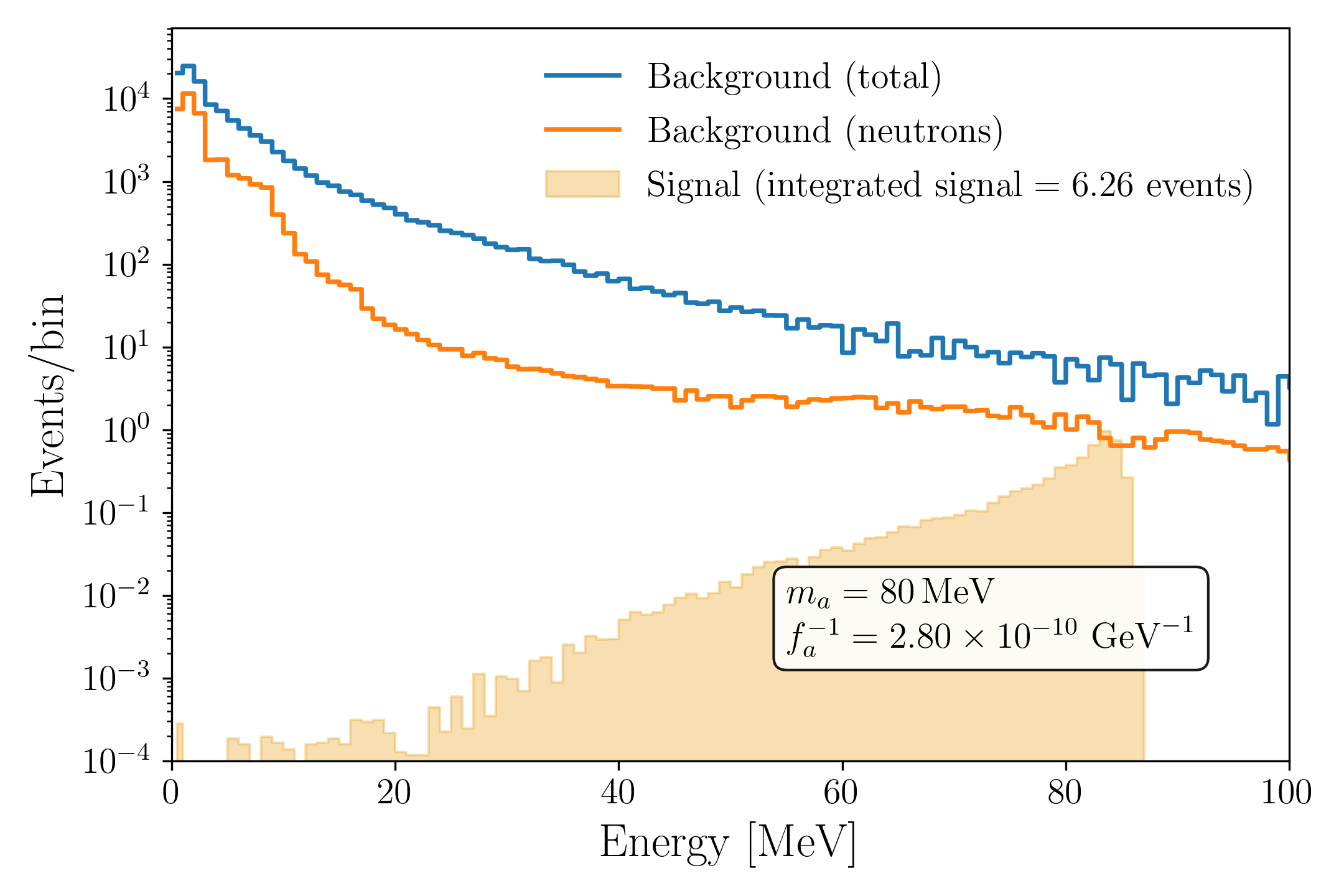}
    
    \caption{Signal vs cosmogenic background event rates in the HC$^2$ detector as a function of the reconstructed energy for HNLs (left column) and LFV ALPs (right column), for coupling values near our sensitivity and for two representative masses (20, 80 MeV). For the background, we also show in orange the contribution from neutrons only, corresponding to the scenario in which the cosmic $\mu^{\pm}$ background could be substantially reduced.
\label{fig:hnl_spectra}
    }
\end{figure*}

\section{Modeling HC$^2$ Response to BSM Decay Products}
\label{sec:response}

To determine the sensitivity of a HC$^2$ campaign at the SNS, we must also describe the response of the detector to the decay $e^+e^-$ signals of interest.  
Response was determined using the set of $e^+e^-$ MC simulations described in Section~\ref{sec:det} that follow the kinematic distributions expected from ALP and HNL decays.  
At each considered point in HNL/ALP parameter space, we determined the total detection efficiency of the signal selection to $e^+e^-$ generated within the HC$^2$ target volume, as well as the distribution of selected events in E$_{rec,c}$ space.

Example efficiencies and reconstructed energy spectra are provided in Figure~\ref{fig:HC_signalMC} for one point in ALP and HNL phase space.  
In the case of 2-body ALP decays, the mono-energetic nature of the $e^+e^-$ pair's summed energy is maintained to a reasonable extent by HC$^2$, with the peak of the signal distribution in Figure~\ref{fig:HC_signalMC} exhibiting a fitted Gaussian $1\sigma$ width of 1.8\%.  
The reconstructed signal distribution from the HNL case is much broader, as expected from the 3-body decay processes involved in obtaining the final-state $e^+e^-$.  
Overall efficiency for the pictured ALP (HNL) case is 12.8\% (16.5\%), with signal losses overwhelmingly dominated by the selection's harsh fiducialization requirements.  
The selection efficiencies for the two cases exhibit some variation across the considered BSM phase space, ranging from approximately 10\% to 17\% for the ALPs and from 15\% to 18\% for the HNLs.  
Lowest efficiencies are observed at 105~MeV, while the highest are observed at lower masses. 

Additionally, for most HNL/ALP masses, only around $\sim 59\%$ of the signal falls within the delayed time window of $1$--$6~\mu\mathrm{s}$. As the masses approach the kinematic threshold, around $80$--$100~\mathrm{MeV}$, this fraction increases up to $\sim 80\%$. For values very close to $m_\mu$, the HNLs/ALPs are produced with near-zero momenta and are no longer visible in the delayed time window. However, this drop in efficiency coincides with the already suppressed production rate due to phase-space closure, and therefore has no visible impact on our final results. For simplicity, we adopt a conservative flat efficiency of $59\%$ for all masses and scenarios.  

\section{Projected \HCdet Sensitivity}
\label{sec:sens}

In this section, we detail our computation of the future sensitivity for our two benchmark scenarios, HNLs and LFV ALPs. 
We use the cosmogenic backgrounds computed in Section~\ref{sec:results2} and the energy response and efficiencies described in Section~\ref{sec:response}.
In Figure~\ref{fig:hnl_spectra}, we compare the number of signal events to the cosmogenic background as a function of the total reconstructed energy ($E_{e^+}+E_{e^-}$, represented by E$_{rec,c}$). 
The left panels correspond to HNLs, while the right panels show LFV ALPs, for two representative masses: 20 and 80 MeV. 
For the background, we show in blue the total cosmogenic contribution and in orange the neutron-induced component. 
For the signal, we choose coupling values that are close to our sensitivity reach when only the neutron background is considered. 
We observe that the HNL signal exhibits a broader energy spectrum, whereas the ALP signal is sharply peaked. 
This difference is due to two body decays in both production and detection of the ALP, leading to a monochromatic line in true energy.

To compute the future sensitivity, we perform a binned Poissonian $\chi^2$ analysis in reconstructed energy.  
The null hypothesis corresponds to the Standard Model expectation (cosmic background only), while the alternative hypothesis includes the contribution from new physics (signal plus background).  
For each reconstructed energy bin $i$, the expected number of events is given by $B_i$ under the null hypothesis and $S_i + B_i$ under the alternative hypothesis, where $S_i$ and $B_i$ denote the predicted signal and background yields, respectively.  
Practically, the background-only expectation of the null hypothesis would be measured with high statistical precision in an HC$^2$ campaign during extended off-beam periods; thus, statistical uncertainties associated with the null hypothesis are not considered.  
The test statistic is constructed from the Poisson likelihood using the Asimov dataset for the nominal HC$^2$ campaign described in  Section~\ref{sec:layout}~\cite{Cowan:2010js}.  
This means that, over the full beam-on-dataset, the observed number of events in each bin $i$, and the statistical uncertainty, is dictated by the expected background, $B_i$.  
Given the low expected signal and backgrounds rates in Table~\ref{tab:dru_HC} and Figure~\ref{fig:hnl_spectra}, we expect statistics-dominated uncertainties and thus will not consider systematic uncertainties related to signal rates or acceptance efficiencies.  
The resulting $\chi^2$ is then given by
\begin{equation}
\chi^2 = 2 \sum_{i } \left[ (S_i + B_i) - B_i + B_i \ln\left(\frac{B_i}{S_i + B_i}\right) \right].
\end{equation}

\begin{figure}[t]
    \includegraphics[width=\columnwidth]{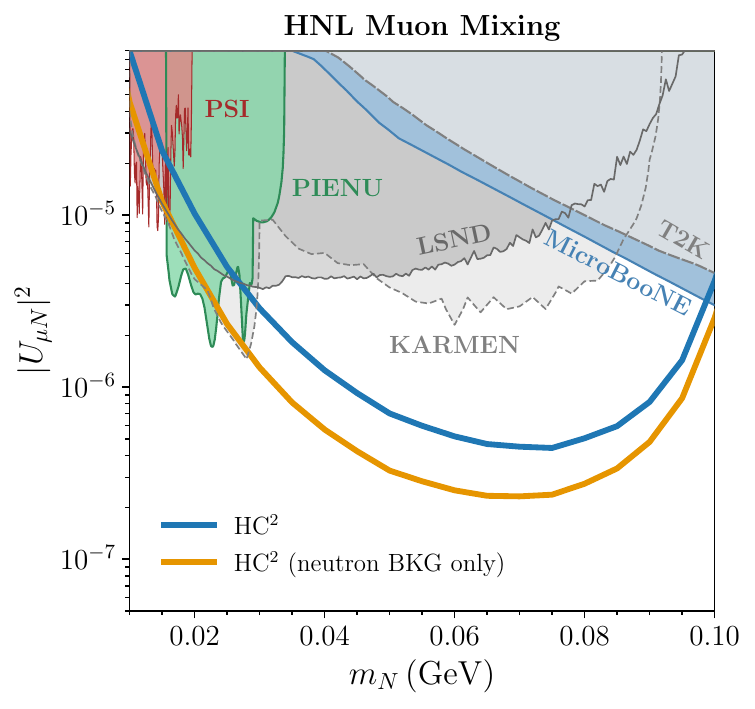}
    \caption{$90\%$ C.L. statistics-only sensitivity curves of the nominal HC$^2$ deployment scenario to heavy neutral leptons with muon mixing for two different background scenarios. 
    The current experimental bounds are taken from Ref.~\cite{Fernandez-Martinez:2023phj} and shown as shaded regions that are colored for dedicated experimental searches or grey for phenomenological recasts.  
    Direct limits from PIENU~\cite{PIENU:2019usb}, MicroBooNE~\cite{MicroBooNE:2023eef} (see also~\cite{Kelly:2021xbv}), and PSI~\cite{Daum:1987bg} are shown in color.
    The constraints from T2K are taken from Ref.~\cite{Arguelles:2021dqn} (a recast of the experimental analysis in \cite{T2K:2019jwa}), while those from LSND and KARMEN are derived from the recast of Refs.~\cite{Hostert:2025ffy} (see also \cite{Ema:2023buz}). 
    \label{fig:HNL_sensitivity}
    }
\end{figure}

\Cref{fig:HNL_sensitivity} shows the resulting 90\% confidence level (C.L.) \HCdet sensitivities for the HNL model\footnote{To set the 90\% C.L. limit, we have assumed the validity of Wilks’ theorem~\cite{Wilks:1938dza}. We have sampled the Poissonian likelihood for some representative masses and found very similar 90\% C.L. regions.
}. 
In this case, we present the results assuming mixing with the muon flavor only. 
The solid lines correspond to two different background assumptions: in blue, we include both cosmogenic backgrounds from muons and neutrons, while in orange we consider only the neutron component, leaving open the possibility of detector additions capable of tagging and rejecting the muon-induced background.  
The neutron-only case should be viewed as a general guideline since realized sensitivity may be somewhat better (\emph{i.e.} due to reductions in $n$-induced cosmics in the shallow-overburden `B' detector location in Figure~\ref{fig:example_fig}) or worse (\emph{i.e.} due to full incorporation of the sub-dominant neutrino background) than for the nominal scenario.  

The shaded colored regions in Figure~\ref{fig:HNL_sensitivity} represent current experimental bounds from peak searches, such as PSI~\cite{Daum:1987bg} and PIENU~\cite{PIENU:2019usb}, as well as the \mbox{MicroBooNE}~\cite{MicroBooNE:2023eef} $N\rightarrow \nu e^+e^-$ search. 
We also show in grey the bounds derived from phenomenological studies, including T2K~\cite{T2K:2019jwa} and recasts from old data of spallation source experiments such as KARMEN and LSND~\cite{Hostert:2025ffy}.
Lower limits on $|U_{\mu N}|^2$ from cosmology also apply in this mass region, effectively constraining the lifetime of the HNL to be less than $\mathcal{O}(0.1$~s$)$~\cite{Sarkar:1995dd,Dolgov:2000jw,Ruchayskiy:2012si,Hufnagel:2017dgo,Sabti:2020yrt,Bondarenko:2021cpc}.
While applicable to the scenario considered here, these can be significantly modified in alternative cosmological scenarios or in non-minimal HNL models~\cite{Arguelles:2021dqn}.

We find that $\text{HC}^2$ has the potential to significantly improve the existing bounds by more than one order of magnitude.
It is worth noting the complementarity of BSM search coverage between measurements at the lower-energy ORNL and higher-energy FNAL proton sources: MicroBooNE limits from the 120~GeV NuMI beam ramp up with increasing HNL mass and decreasing boost factors, only achieving best capabilities after HC$^2$ sensitivity tails off moving upwards towards the muon mass.
Above the muon mass, higher-energy beams can constrain HNLs produced in heavier mesons that are not produced at the SNS.

\begin{figure}[htpb!]
    \includegraphics[width=0.5\textwidth]{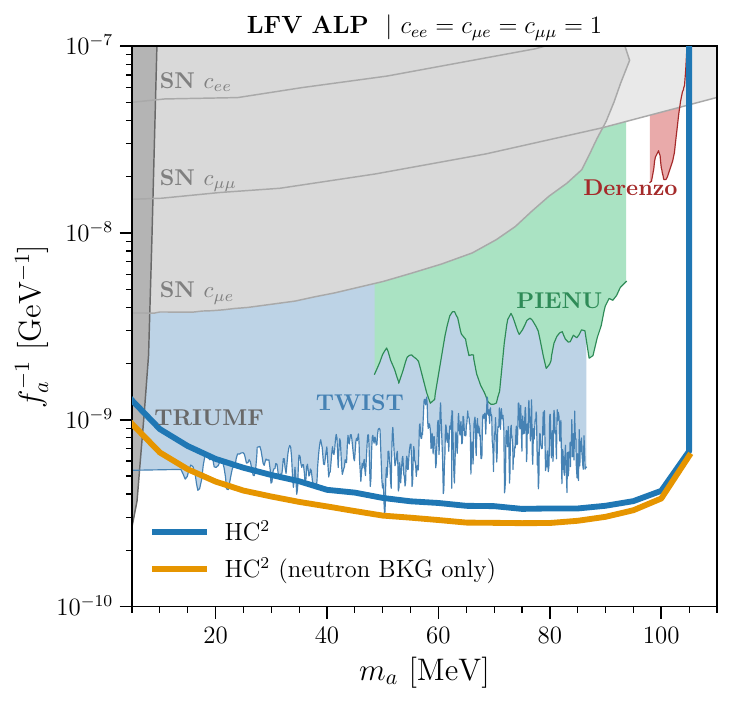}
    \includegraphics[width=0.5\textwidth]{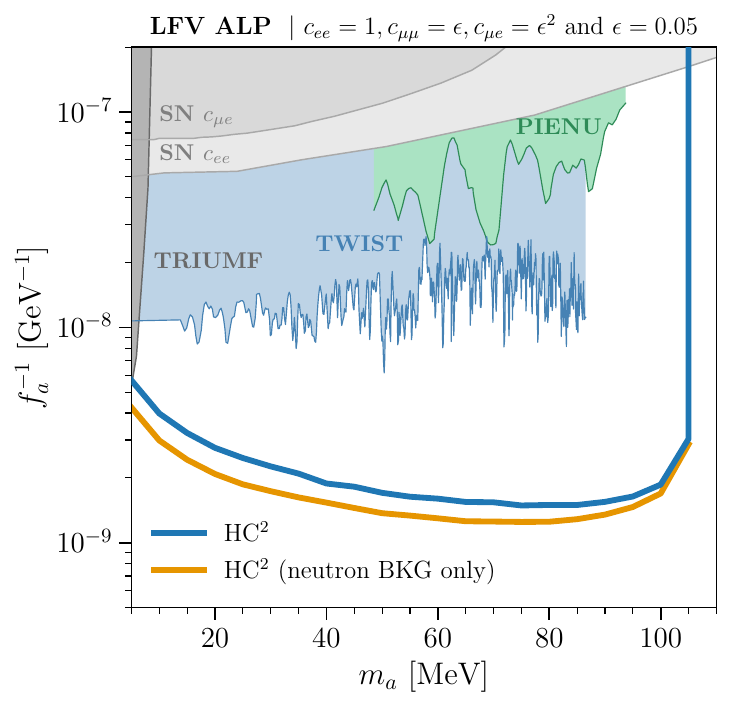}
    \caption{
    $90\%$ C.L. statistics-only sensitivity curves for the nominal HC$^2$ deployment scenario in the ALP parameter space for two different background scenarios (blue and orange solid lines) and two choices of ALP couplings to leptons.
    The top panel shows the case where the ALP couples equally to all leptons in the derivative basis.
    The bottom panel shows the case where the ALP couples more strongly to flavor-diagonal currents. 
    Also shown are direct bounds from muon decay experiments TWIST~\cite{TWIST:2014ymv}, Derenzo~\cite{Derenzo:1969za}, and PIENU~\cite{PIENU:2020loi}.
    Phenomenological bounds are shown in shades of grey, including a recast~\cite{Calibbi:2020jvd} of a search for right-handed currents in muon decay at TRIUMF~\cite{Jodidio:1986mz} and supernova bounds on individual scenarios with flavor-diagonal and off-diagonal couplings~\cite{Li:2025beu,Huang:2025rmy,Carenza:2021pcm,Fiorillo:2025sln} (see Refs.~\cite{Calibbi:2020jvd,Hostert:2025ffy} for a discussion).
    }
    \label{fig:ALP_sensitivity}
\end{figure}

\Cref{fig:ALP_sensitivity} shows our results for the LFV ALP scenario.
In this model, the best existing bounds in the relevant mass range arise from searches for the two-body decay of a muon to an electron and an invisible boson, ($\mu \to e X_{\rm inv}$) at TWIST~\cite{TWIST:2014ymv}, PSI~\cite{PIENU:2020loi}, and earlier facilities~\cite{Derenzo:1969za,Jodidio:1986mz}, together with supernova constraints~\cite{Li:2025beu}. 
These searches are particularly powerful because the two-body decay benefits from the enhancement of the muon decay width proportional to $1/(f^2_aG_F^2 m_\mu^2)$, allowing them to probe decay constants as large as $f_a \sim 10^9$~GeV.

In the upper panel of \cref{fig:ALP_sensitivity}, we consider the case where the flavor-violating and flavor-conserving couplings are of the same size (in the derivative basis of \cref{eq:ALPcurrent}). 
For masses below approximately 80 MeV, existing $\mu \rightarrow e X_{\text{inv}}$ searches already impose strong constraints, and our setup improves these limits only moderately. 
However, as the ALP mass approaches the kinematic endpoint of the muon decay, the monochromatic electron signal in muon experiments becomes increasingly lower-energy and harder to identify on top of the dominant muon decay branching ratio, causing the sensitivity of peak-search experiments to deteriorate rapidly. 
By contrast, the decay-in-flight strategy remains essentially stable in this region, providing a clear advantage near the kinematic threshold, where the sensitivity improves by nearly two orders of magnitude.

By varying $c_{\mu e}$ independently from $c_{ee}$, we can effectively decouple the branching ratio $\mu \rightarrow e X_{\text{inv}}$, proportional to $c_{\mu e}$, from the ALP decay rates $\Gamma(a\to e^+e^-)$, proportional to $c_{ee}$.
Our decay-in-flight strategy is sensitive to their product, while precision muon experiments are sensitive to the muon branching ratio only, probing complementary parameter space.
A favorable combination of couplings for decay-in-flight searches includes a hierarchy between flavor-violating and flavor-conserving couplings.
As a representative point, we show the case where $c_{\mu e}/c_{ee} = \epsilon^2 = 0.0025$ in the lower panel of \cref{fig:ALP_sensitivity}. 
The muon decay experiments can only constrain the off-diagonal coupling $c_{\mu e}$ and are only sensitive to $f_a \sim 10^{8}$~GeV for our choice of parameters, being mostly insensitive to the diagonal coupling $c_{ee}$, provided the ALP decays outside the detector, which is indeed the case throughout the open parameter space. 
As a result, the decay-in-flight strategy offers clear advantage when $c_{\mu e} \ll c_{ee}$.
The ratio between decay-in-flight limits and muon experiment limits on $f_a$ scale approximately as $f_a^{\mu \to e X_{\rm inv}}/f_a^{\rm DIF} = \sqrt{c_{\mu e}/c_{ee}}$ in the long lifetime regime.

\section{Comments on Neutrino Interaction Measurements}\label{sec:neutrino}

This study has demonstrated that cosmic backgrounds to leptonic BSM particle decay signatures from the SNS target could be expected to yield $\mathcal{O}$(hundreds) of counts or fewer over the course of a 3-year campaign with a well-designed HC$^2$ detector implementation.  
This outcome leads us to consider what rate of SNS-induced leptonic \emph{neutrino} signals might be recorded during this time.  
In particular, the inelastic charged-current (CC) interaction $\nu_e + ^{12}$C$\rightarrow ^{12}$N$+e^-$, occurring in the target scintillator region, will generate final-state leptons with energies of up to 40~MeV~\cite{Formaggio:2012cpf} with broad spectra  somewhat comparable to those pictured in~Figure~\ref{fig:signal_at_detector} for final-state $e^+e^-$ pairs from the lowest-mass HNL case.  
Interactions generating ground-state $^{12}$N (`CNgs') are accompanied by a delayed $\beta^+$-decay signature (t$_{1/2}$=11~ms; Q$_{\beta^+}$=16.3~MeV), while excited-state $^{12}$N$^*$ products are not, since they decay promptly via proton emission.  
Both of these interaction channels have been previously measured by the LSND~\cite{LSND:1997lta,LSND:2001fbw} and KARMEN experiments~\cite{KARMEN:1994xse}, with (10-15)\% errors dominated by uncertainties in the neutrino flux emitted by their respective beam facilities~\cite{Formaggio:2012cpf}.  

An exclusive measurement of CNgs recently reported by the JSNS$^2$ collaboration~\cite{JSNS2:2024uxo} provides an excellent foundation for estimating expected rates in a 3-year HC$^2$ deployment.  
Over 2.2$\times$10$^{22}$ POT of data-taking at 24~m from the 3 GeV MLF beam's target, JSNS$^2$ detected 37 CNgs events in 10 fiducial tons of LS, with selection cuts achieving 5.88\% efficiency for all CNgs events in the fiducial volume.  
In comparison, the nominal HC$^2$ deployment involves a 4~ton target performing 5$\times$10$^{23}$ POT of data-taking at 20~m from the 1.3 GeV SNS beam target.  
Scaling based on these relative parameter differences, and assuming a HC$^2$ detection efficiency roughly matching the $\sim$15\% achieved at the lowest HNL energies, one should expect to measure hundreds of events across the two $^{12}$C CC interaction channels.  
Only the `single-cluster' $^{12}$N$^*$ channel would serve as a potential background to ALP and HNL measurements, providing a potential sub-dominant contribution to cosmic backgrounds for the described potential HC$^2$ implementations.

If, however, a HC$^2$ implementation can be accompanied by modest overburden and a muon veto sub-system, neutrino-dominated measurements may potentially be achievable for both CC interaction channels on $^{12}$C.  
Further, if COHERENT's D2O detector modules achieve their goal of constraining SNS neutrino flux uncertainties to the few-percent level~\cite{COHERENT:2021xhx}, HC$^2$ neutrino measurements could achieve leading precision in measurements of $^{12}$C interactions.  
Increased precision (particularly of the more experimentally and theoretically uncertain $^{12}$N$^*$ channel) can be used to improve low-energy neutrino interaction generators~\cite{Gardiner:2021qfr} and strengthen predictions for future astrophysical neutrino measurements for large hydrocarbon neutrino detectors, such as JUNO~\cite{JUNO:2021vlw,JUNO:2022lpc,JUNO:2023dnp}.  

While rates of elastic neutrino-nucleus scattering on $^1$H within a HC$^2$ target are expected to be far too low to perform measurement of this detection channel, the endpoint of its $p$-recoil signature, appearing in the ~1~MeV regime, is theoretically within the dynamic range of a HC$^2$ detector.  
Thus, a HC$^2$ detector could also potentially offer sensitivity to non-standard neutrino interaction effects manifesting themselves in elastic or coherent neutrino-nucleus scattering (CE$\nu$NS) interactions~\cite{Giunti:2019xpr}.  
Due to the large amount of scintillation quenching experienced by heavy nuclear recoils, we do not expect CE$\nu$NS on $^{12}$C to be visible above backgrounds in a HC$^2$ detector.  
Finally, high-energy ($\sim$10-20~MeV) $\gamma$-rays could be generated in HC$^2$ via neutral current $\nu_e$- $^{12}$C scattering (E$_{\gamma}$ = 15.1~MeV)~\cite{KARMEN:1991vkr} or via inelastic scattering of SNS-produced dark matter~\cite{Dutta:2024yjp,Brdar:2025hqi,Dutta:2026zpe,Dutta:2026dvg}.  
While the signature of these $\gamma$-rays should be detectable, they lie outside the signal region studied in this paper, so we refrain from further comment here.

\section{Summary}
\label{sec:conc}

In this paper, we have demonstrated and discussed exciting LLP search possibilities offered by the deployment of a tons-scale highly capable hydrocarbon scintillator detector at the SNS.  
The HC$^2$ concept we present is distinct from recent spallation-source hydrocarbon detector deployments, like JSNS$^2$, in its focus on time-isolated, rather than time-coincident, physics signatures.  
In this sense, the HC$^2$ concept pursues multi-MeV scale physics signatures more akin to those targeted by non-hydrogenous COHERENT sub-detectors, such as LAr-750, NaIvETe, or D20~\cite{COHERENT:2026ewu}, but with a broader arsenal of background rejection capabilities at its disposal.  

We showcased the utility of this background rejection arsenal by performing a search for isolated $e^+e^-$ signatures in the on-surface PROSPECT detector during reactor-off periods.  
The signal selection was phenomenologically informed via calculation and simulation of $e^+e^-$ signatures expected from HNL and ALP muon decay products from the SNS target.  
This search results in roughly 1 cosmic background count for every 8 seconds of live time — roughly 1/2 day’s worth of post-beam-spill signal windows at the SNS — in the 4-ton PROSPECT target.  
Good signal rate and energy spectrum agreement are found between PROSPECT’s cosmic data and MC simulations.  

We then simulated cosmic backgrounds in a range of potential 4~ton HC$^2$ detector implementations at the SNS, finding that a segmented, PSD-capable, un-doped HC$^2$ detector, combined with an external muon veto, may be able to achieve few-hundred-level background counts for HNL/ALP-like signals in a three-year SNS data-taking campaign.  
Such a campaign could achieve world-leading coverage of ALP and HNL phase space in the 10-100 MeV mass regime, in some cases beating out existing limits by multiple orders of magnitude.  
Sensitivity is conditional on the impact of uncharacterized neutron-induced $\gamma$-ray backgrounds (such as activation and neutron capture) on analysis cut efficiencies; thus, near-term radiation surveys at the two studied SNS deployment locations would be important for establishing feasibility and informing passive shielding design requirements.  
It also seems possible that this same HC$^2$ campaign could measure inelastic neutrino scattering on $^{12}$C with precision comparable to prior state-of-the-art measurements from LSND and KARMEN, while also pursuing sensitivity to other new physics models via searches for CE$\nu$NS-like signatures on $^1$H or $>$10~MeV nuclear de-excitation $\gamma$-rays.  

\begin{acknowledgments}

We acknowledge and thank Dan Pershey for helpful discussions regarding $\nu_e$ inelastic scattering at the SNS.  

This material is based upon work supported by the following sources: US Department of Energy (DOE) Office of Science, Office of High Energy Physics under Award No. DE-SC0016357 and DE-SC0017660 to Yale University, under Award No. DE-SC0017815 to Drexel University, under Award No. DE-SC0008347 to Illinois Institute of Technology, under Award No. DE-SC0010504 to University of Hawaii, under Contract No. DE-SC0012704 to Brookhaven National Laboratory, and under Work Proposal Number  SCW1504 to Lawrence Livermore National Laboratory.  
This work was performed under the auspices of the U.S. Department of Energy by Lawrence Livermore National Laboratory under Contract DE-AC52-07NA27344 and by Oak Ridge National Laboratory under Contract DE-AC05-00OR22725.  
Additional funding for the experiment was provided by the Heising-Simons Foundation under Award No. \#2016-117 to Yale University.  
M.H. was partially supported by the University of Iowa’s Year 2 P3 Strategic Initiatives Program through funding received for the project entitled ``High Impact Hiring Initiative (HIHI): A Program to Strategically Recruit and Retain Talented Faculty.''

 
We further acknowledge support from Yale University, the Illinois Institute of Technology, Temple University, University of Hawaii, Brookhaven National Laboratory, the Lawrence Livermore National Laboratory LDRD program, the National Institute of Standards and Technology, and Oak Ridge National Laboratory. We gratefully acknowledge the support and hospitality of the High Flux Isotope Reactor and Oak Ridge National Laboratory, managed by UT-Battelle for the U.S. Department of Energy.  The views expressed in this paper are those of the authors and do not reflect the official policy or position of the U.S. Naval Academy, Department of the Navy, Department of Defense, or U.S. Government.
\end{acknowledgments}

\section*{Data Availability}
Data and analysis code that support the findings of this study are available at \href{https://github.com/SalvaUrrea2/HC2_study}{GitHub repository} upon publication.

\bibliographystyle{apsrev4-2}
\bibliography{refs}

\end{document}

%% file: AuthorListJune2026.tex
\affiliation{Department of Physics, Boston University, Boston, Massachusetts, USA}
\affiliation{Brookhaven National Laboratory, Upton, New York, USA}

\affiliation{Department of Physics, Drexel University, Philadelphia, Pennsylvania, USA}
\affiliation{George W.\,Woodruff School of Mechanical Engineering, Georgia Institute of Technology, Atlanta, Georgia, USA}
\affiliation{Department of Physics \& Astronomy, University of Hawaii, Honolulu, Hawaii, USA}
\affiliation{Department of Physics, Illinois Institute of Technology, Chicago, Illinois, USA}
\affiliation{Department of Physics and Astronomy, Johns Hopkins University, Baltimore, Maryland, USA}
\affiliation{Nuclear and Chemical Sciences Division, Lawrence Livermore National Laboratory, Livermore, California, USA}
\affiliation{Department of Physics, Le Moyne College, Syracuse, New York, USA}
\affiliation{National Institute of Standards and Technology, Gaithersburg, Maryland, USA}
\affiliation{High Flux Isotope Reactor, Oak Ridge National Laboratory, Oak Ridge, Tennessee, USA}
\affiliation{Physics Division, Oak Ridge National Laboratory, Oak Ridge, Tennessee, USA}
\affiliation{Department of Physics, Susquehanna University, Selinsgrove, Pennsylvania, USA}
\affiliation{Department of Physics and Astronomy, University of Tennessee, Knoxville, Tennessee, USA}
\affiliation{Department of Physics, University of Wisconsin, Madison, Wisconsin, USA}
\affiliation{Wright Laboratory, Department of Physics, Yale University, New Haven, Connecticut, USA}

\author{M.\,Andriamirado}\affiliation{Department of Physics, Illinois Institute of Technology, Chicago, Illinois, USA}
\author{A.\,B.\,Balantekin}\affiliation{Department of Physics, University of Wisconsin, Madison, Wisconsin, USA}
\author{C.\,D.\,Bass}\affiliation{Department of Physics, Le Moyne College, Syracuse, New York, USA}
\author{O.\,Benevides Rodrigues}\affiliation{Department of Physics, Illinois Institute of Technology, Chicago, Illinois, USA}
\author{E.\,P.\,Bernard}\affiliation{Nuclear and Chemical Sciences Division, Lawrence Livermore National Laboratory, Livermore, California, USA}
\author{N.\,S.\,Bowden}\affiliation{Nuclear and Chemical Sciences Division, Lawrence Livermore National Laboratory, Livermore, California, USA}
\author{C.\,D.\,Bryan}\affiliation{High Flux Isotope Reactor, Oak Ridge National Laboratory, Oak Ridge, Tennessee, USA}
\author{R.\,Carr}\affiliation{United States Naval Academy, Annapolis, MD, USA}
\author{T.\,Classen}\affiliation{Nuclear and Chemical Sciences Division, Lawrence Livermore National Laboratory, Livermore, California, USA}
\author{A.\,J.\,Conant}\affiliation{High Flux Isotope Reactor, Oak Ridge National Laboratory, Oak Ridge, Tennessee, USA}
\author{N.\,Craft}\affiliation{Department of Physics, Drexel University, Philadelphia, Pennsylvania, USA}
\author{G.\,Deichert}\affiliation{High Flux Isotope Reactor, Oak Ridge National Laboratory, Oak Ridge, Tennessee, USA}
\author{A.\,Erickson}\affiliation{George W.\,Woodruff School of Mechanical Engineering, Georgia Institute of Technology, Atlanta, Georgia, USA}
\author{M.\,D.\,Fuller}\affiliation{High Flux Isotope Reactor, Oak Ridge National Laboratory, Oak Ridge, Tennessee, USA}
\author{A.\,Galindo-Uribarri}\affiliation{Physics Division, Oak Ridge National Laboratory, Oak Ridge, Tennessee, USA} \affiliation{Department of Physics and Astronomy, University of Tennessee, Knoxville, Tennessee, USA}
\author{S.\,Ghosh}\affiliation{Nuclear and Chemical Sciences Division, Lawrence Livermore National Laboratory, Livermore, California, USA}
\author{S.\,Gokhale}\affiliation{Brookhaven National Laboratory, Upton, New York, USA}
\author{C.\,Grant}\affiliation{Department of Physics, Boston University, Boston, Massachusetts, USA}
\author{S.\,Hans}\affiliation{Brookhaven National Laboratory, Upton, New York, USA}
\author{A.\,B.\,Hansell}\affiliation{Department of Physics, Susquehanna University, Selinsgrove, Pennsylvania, USA}
\author{T.\,E.\,Haugen}\affiliation{National Institute of Standards and Technology, Gaithersburg, Maryland, USA}
\author{K.\,M.\,Heeger}\affiliation{Wright Laboratory, Department of Physics, Yale University, New Haven, Connecticut, USA}
\author{A.\,Irani}\affiliation{Department of Physics, Illinois Institute of Technology, Chicago, Illinois, USA}
\author{J. Koblanski}\affiliation{Department of Physics \& Astronomy, University of Hawaii, Honolulu, Hawaii, USA}
\author{C.\,E.\,Lane}\affiliation{Department of Physics, Drexel University, Philadelphia, Pennsylvania, USA}
\author{B.\,R.\,Littlejohn}\affiliation{Department of Physics, Illinois Institute of Technology, Chicago, Illinois, USA}
\author{A.\,Lozano Sanchez}\affiliation{Department of Physics, Drexel University, Philadelphia, Pennsylvania, USA}
\author{F.\,Machado}\affiliation{Department of Physics, Illinois Institute of Technology, Chicago, Illinois, USA}
\author{J.\,Maricic}\affiliation{Department of Physics \& Astronomy, University of Hawaii, Honolulu, Hawaii, USA}
\author{M.\,P.\,Mendenhall}\affiliation{Nuclear and Chemical Sciences Division, Lawrence Livermore National Laboratory, Livermore, California, USA}
\author{A.\,M.\,Meyer}\affiliation{Department of Physics \& Astronomy, University of Hawaii, Honolulu, Hawaii, USA}
\author{R.\,Milincic}\affiliation{Department of Physics \& Astronomy, University of Hawaii, Honolulu, Hawaii, USA}
\author{P.\,E.\,Mueller}\affiliation{Physics Division, Oak Ridge National Laboratory, Oak Ridge, Tennessee, USA} 
\author{H.\,P.\,Mumm}\affiliation{National Institute of Standards and Technology, Gaithersburg, Maryland, USA}
\author{R.\,Neilson}\affiliation{Department of Physics, Drexel University, Philadelphia, Pennsylvania, USA}
\author{J.\,R.\,Newby}\affiliation{Physics Division, Oak Ridge National Laboratory, Oak Ridge, Tennessee, USA} 
\author{D.\,Norcini}\affiliation{Department of Physics and Astronomy, Johns Hopkins University, Baltimore, Maryland, USA} 
\author{N.\,Patel}\affiliation{Department of Physics \& Astronomy, University of Hawaii, Honolulu, Hawaii, USA}
\author{C.\,Roca}\affiliation{Nuclear and Chemical Sciences Division, Lawrence Livermore National Laboratory, Livermore, California, USA}
\author{R.\,Rosero}\affiliation{Brookhaven National Laboratory, Upton, New York, USA}
\author{D.\,Venegas-Vargas}\affiliation{Department of Physics and Astronomy, Johns Hopkins University, Baltimore, Maryland, USA}
\author{J.\,Wilhelmi}\affiliation{Wright Laboratory, Department of Physics, Yale University, New Haven, Connecticut, USA}
\author{M.\,Yeh}\affiliation{Brookhaven National Laboratory, Upton, New York, USA}
\author{X.\,Zhang}\affiliation{Nuclear and Chemical Sciences Division, Lawrence Livermore National Laboratory, Livermore, California, USA}